\documentclass[11pt,twoside,a4paper]{article}

\pdfoutput=1

\usepackage{amsmath}
\usepackage{amssymb}
\usepackage{epsfig}
\usepackage{graphicx}
\usepackage{psfrag}
\usepackage[english]{babel}
\usepackage{fontenc}
\usepackage[latin9]{inputenc}
\usepackage{xspace}
\usepackage[left=1.5cm,right=1.5cm,top=2.5cm,bottom=2.5cm]{geometry}
\usepackage{color}
\usepackage{cite}
\definecolor{vertp}{rgb}{0,1,0}
\definecolor{rouge}{rgb}{1,0,0}
\definecolor{bleu}{rgb}{0,0,1}
\definecolor{rose}{rgb}{1,0,0.5}
\definecolor{orange}{rgb}{1,0.5,0}
\definecolor{bleuf}{rgb}{0.2,0.2,0.5}
\definecolor{cyan}{rgb}{0.5,1,1}
\definecolor{vertg}{rgb}{0.5,1,0}
\definecolor{magenta}{rgb}{1,0,1}
\definecolor{mauve}{rgb}{0.5,0,0.5}

\def\gsim{\mathrel{\lower3pt\hbox{$\sim$}}\hskip-11.5pt\raise3pt\hbox{$>$}\;}

\def\lsim{\mathrel{\lower3pt\hbox{$\sim$}}\hskip-11.5pt\raise3pt\hbox{$<$}\;}

\title{}
\date{}

\begin{document}
\vspace{1.cm}
\flushright{ULB-TH/08-40}
\vspace{1.cm}

\begin{center}

{\Large {\bf  $e^+$ and $\bar p$ from inert doublet model dark matter}}
\vspace{0.8cm}\\
{\large  Emmanuel Nezri$^1$, Michel H.G. Tytgat$^2$ and Gilles Vertongen$^2$}
\vspace{0.35cm}\\
$^1${\it 
Laboratoire d'Astrophysique de Marseille,
OAMP,\\
Technop\^ole de Marseille-Etoile. 38, rue Fr\'ed\'eric Joliot-Curie 13388
Marseille, France}
\vspace{0.35cm}\\
$^2${\it 
Service de Physique Th\'eorique, Universit\'e Libre de Bruxelles,\\
CP225, Bld du Triomphe, 1050 Brussels, Belgium}
\vspace{0.5cm}\\

\end{center}
\abstract{       
In the framework of the Inert Doublet Model, a very simple extension of the Standard Model, we study the production and propagation of antimatter in cosmic rays coming from annihilation of a scalar dark matter particle. 
We consider  three benchmark candidates, all consistent with the WMAP cosmic abundance and existing direct detection experiments, and confront the predictions of the model with the recent PAMELA, ATIC and HESS data. For a light candidate, $M_{DM} \sim 10$ GeV, we argue that the positron and anti-proton fluxes may be large, but still consistent with expected backgrounds, unless there is an enhancement (boost factor) in the local density of dark matter. There is also a substantial anti-deuteron flux which might be observable by future experiments. For a candidate with $M_{DM} \sim 70$ GeV, the contribution to $e^+$ and $\bar p$ fluxes is much smaller than the expected backgrounds. Even if a boost factor is invoked to enhance the signals, the candidate is unable to explain the observed $e^+$ and $\bar p$ excesses. Finally, for a heavy candidate, $M_{DM} \sim 10 $\,TeV, it is possible to fit the PAMELA excess (but, unfortunately, not the ATIC one) provided there is a large enhancement, either in the local density of dark matter or through the Sommerfeld effect.
}

\section{Introduction}

The recent cosmological observations indicate that about 80\% of matter in the universe is made of dark matter (DM) \cite{Komatsu:2008hk,Seljak:2006bg,Yao:2006px}. 
The most popular, particle physics, explanation is that DM is made of weakly interacting massive particles, or WIMPs \cite{Jungman:1995df,Bertone:2004pz}, and the leading WIMP candidate is the neutralino, a supersymmetric spin $1/2$ Majorana particle. However, spin one WIMPs, like in models with extra dimensions \cite{Hooper:2007qk}, and, to a lesser extent,  scalar WIMPs, are considered
as interesting challengers.  
An instance of the latter is the lightest stable scalar of the Inert Doublet Model (IDM), an extension of the Standard Model with two Higgs doublets  and a discrete $Z_2$ symmetry which is imposed as a simple way to prevent flavour changing neutral currents (FCNC)\cite{Deshpande:1977rw}. The scope and ambition of the IDM can not compete with those of the MSSM or models with extra dimensions, but the IDM is a simple model and, nevertheless, it does have
an  interesting phenomenology, as emphasised in \cite{Ma:2006km,Barbieri:2006dq}, and {\em e.g.} \cite{Hambye:2007vf}. Also, it encompasses some of the features of other models with scalar dark matter, like Minimal Dark Matter scalar candidates \cite{Cirelli:2005uq}, or  singlet scalars, like in hidden portal models \cite{McDonald:1993ex,Patt:2006fw}.

The prospects for the direct and indirect detection of IDM dark matter, both through gamma rays from the Galactic Centre (GC), and through neutrinos from the Earth and Sun, have been addressed in \cite{Barbieri:2006dq,LopezHonorez:2006gr,Gustafsson:2007pc,Majumdar:2006nt,Andreas:2008xy,Agrawal:2008xz,Andreas:2009hj}. In the present article, 
we study the production and propagation of
positrons ($e^+$) and anti-protons ($\bar p$) that would result from its annihilation in the halo of the Galaxy \footnote{For a summary on the search of dark matter,
including the IDM candidate, see \cite{Taoso:2007qk}.}. The subject
matter is timely, given the recent release of various observations of antimatter in cosmic rays, in particular from the {\em Payload for
Antimatter Exploration and Light-Nuclei Astrophysics} (PAMELA) experiment, which
has  an excess in the positron fraction, $e^+/(e^+ + e^-)$ in the $10-80$ GeV range
\cite{Adriani:2008zr}  (but no excess in the $\bar p/p$ flux \cite{Adriani:2008zq}), the {\em Advanced Thin
Ionization Calorimeter} (ATIC) experiment, which shows an excess of cosmic ray
electrons (plus positrons) at energies of $300-800$ GeV \cite{:2008zz}, and, finally, the {\em High Energy Stereoscopic System} (HESS) experiment, which has measured the flux of cosmic ray electrons (plus positrons) between $0.6-4$ TeV\cite{Collaboration:2008aaa}.

These data, and their puzzling features, have generated a flurry of activity.  As shown in various works \cite{Aharonian:1995he,Hooper:2008kg,Yuksel:2008rf,Profumo:2008ms,Shaviv:2009bu}, the PAMELA and ATIC anomalies may be of astrophysical origin, a natural source of positrons being nearby pulsars. Nevertheless, much of the activity has been focused on the possibility to interpret the excess of positrons as being due to annihilation or decay of dark matter particles in the galactic halo \cite{Cirelli:2008pk,Barger:2008su,Donato:2008jk,Cholis:2008wq,Bertone:2008xr,Cirelli:2008jk,Bergstrom:2008gr,ArkaniHamed:2008qn,Nardi:2008ix,Fox:2008kb,Yin:2008bs}. 

As we have already emphasised, the main advantage of the IDM is it simplicity. This is both a blessing and a curse. 
On one hand, it explains why the IDM should not pretend to
fully explain the recent data. On the other hand, there is no analysis yet of
antimatter production in the IDM, and we see this as an opportunity to fill a gap in the literature.

The paper is organised as follow. The Inert Doublet Model is summarised in Section 2. Anti-proton and positron production and propagation are briefly presented in Section 3, together with a discussion on backgrounds expectations. Section 4 is
devoted to the contribution to cosmic rays of annihilation of a IDM dark matter candidate. Cosmic
rays features specific of the Inert Doublet Model and, in particular, their relevance for the results of PAMELA,  ATIC and HESS experiments,  are
presented in Section 5. We give our conclusions in Section 6.

\section{The Inert Doublet Model}

The Inert Doublet Model is a two Higgs doublet model, noted $H_1$ and $H_2$,  with an unbroken 
$Z_2$ 
symmetry, such that 
$$ 
H_1 \rightarrow H_1 \;\; \mbox{\rm and}\;\; 
H_2\rightarrow - H_2, 
$$
and with all the other Standard Model particles even
\cite{Ma:2006km,Barbieri:2006dq}. 
The potential of the IDM extension of the Standard Model
may be written as 
\begin{equation} 
\label{potential} 
V = \mu_1^2 \vert H_1\vert^2 + \mu_2^2 \vert H_2\vert^2  + \lambda_1 \vert H_1\vert^4 + 
\lambda_2 \vert H_2\vert^4 + \lambda_3 \vert H_1\vert^2 \vert H_2 \vert^2 
+ \lambda_4 \vert H_1^\dagger H_2\vert^2 + \frac{\lambda_5}{2} \left[(H_1^\dagger H_2)^2 + h.c.\right]\,. 
\end{equation} 
In this model, $H_1$ contains the standard Brout-Englert-Higgs particle $h$ (the Higgs for short in the sequel). The discrete symmetry, which prevents FCNC, is assumed to be unbroken, so that the $H_2$ has no vacuum expectation value, and, consequently, the lightest $Z_2$ odd particle is stable. The  dark matter candidate may be either one of the neutral components of the doublet $H_2 = (H^+, 1/\sqrt{2} (H_0 +i A_0))^T$
\begin{eqnarray} 
M_h^2 &=& - 2 \mu_1^2 \equiv 2 \lambda_1 v^2\,, \nonumber\\
M_{H^+}^2 &=& \mu_2^2 + \lambda_3 v^2/2\,,\nonumber\\
M_{H_0}^2 &=& \mu_2^2 +  (\lambda_3 + \lambda_4 + \lambda_5) v^2/2\,,\nonumber\\
M_{A_0}^2 &=& \mu_2^2 +  (\lambda_3 + \lambda_4 - \lambda_5) v^2/2\,.
\label{masses} 
\end{eqnarray}
Depending on quartic couplings, either $H_0$ or $A_0$ can be the lightest particle. We choose $H_0$ and, following \cite{Barbieri:2006dq}, we define
$\lambda_L=(\lambda_3 + \lambda_4 + \lambda_5)/2$, 
which measure the trilinear coupling between  the Higgs $h$,  and a pair of $H_0$. Choosing $A_0$ instead would not change our conclusions. In our investigation of the model we choose $\mu_2$, $\lambda_2$,
and the masses of  scalar particles, including the mass of the Higgs, as input parameters.

Experimental constraints on the IDM model from colliders have been discussed in \cite{Barbieri:2006dq}, and further in \cite{Lundstrom:2008ai}, and  \cite{Cao:2007rm}. The latter also discusses the prospect for discovery of the $A_0$ and $H_0$ at the LHC. In  \cite{Lundstrom:2008ai}, the LEP I and II constraints on the neutralino are used to put constraints on the mass range of both $H_0$ and $A_0$. These constraints are essentially summarised in the Figure 8 of that reference. The three benchmark dark matter candidates we will consider here, as well as the properties of the their companions $A_0$ and $H^\pm$, are all chosen to be consistent with LEP and WMAP data. 

Assuming that the $H_0$ was in thermal equilibrium in the early universe, there are essentially three distinct mass ranges within which an IDM candidate has an abundance consistent with WMAP.  In the sequel we refer to these ranges as the {\em low mass} ($3$ GeV $\lsim M_{H_0} \lsim 8 $\,GeV)\cite{Andreas:2008xy}, the {\em middle mass}  ($40$ GeV $\lsim M_{H_0} \lsim 80$ GeV)\cite{Barbieri:2006dq}, and the {\em high mass} ($500$ GeV $\lsim M_{H_0} \lsim 15$ TeV) ranges\cite{LopezHonorez:2006gr}. 

Candidates in the low mass range may annihilate only through the Higgs. The option of a light $H_0$ is challenged, but not excluded, by current direct detection experiments (CDMS, Xenon and Cogent), and may even be compatible with the DAMA results\cite{Andreas:2008xy}. Candidates in the middle mass range may annihilate through the $Z$ (when co-annihilation with $A_0$ is allowed) or through the Higgs. Some solutions in the middle mass range  may also have a large annihilation cross section, through loop corrections, into a pair of gamma rays. This is relevant for indirect detection from annihilation at the GC, and such a gamma ray line would be easily observed by the GLAST/Fermi satellite \cite{Gustafsson:2007pc}. Above $M_{H_0} \sim 80$ GeV, annihilation into $W^\pm$ pairs is allowed with a large cross section, and the abundance from freeze-out is predicted to be well below the WMAP abundance \cite{LopezHonorez:2006gr}. At even higher masses, however, the annihilation cross section tends to decrease, $\sigma v \propto 1/M_{H_0}^2$, and, consequently, the relic abundance increases, and, for some masses, may be consistent with WMAP observations. In this regime, an IDM dark matter candidate is analogous to that of Minimal Dark Matter \cite{Cirelli:2005uq}. However, in the IDM, there are more parameters to play with, and there is a whole range of DM candidates  around 1 TeV  with a cosmic abundance that is consistent with WMAP observations\cite{LopezHonorez:2006gr}.  

\section{Antimatter in Cosmic Rays}

\subsection{Anti-protons}
The anti-protons in cosmic rays are essentially by-products (secondaries). They  are created by spallation, through 
collisions between protons, which are primary constituents of cosmic rays, and nuclei (essentially hydrogen) from the interstellar medium (ISM). 
The creation, destruction and propagation of anti-protons in the ISM and in the galactic magnetic field, may be described by a diffusion equation for the anti-proton density $n_{\bar{p}}$. Solving this equation requires the distribution of matter in the Galaxy. Its shape is modelled by two cylindrical slabs, one representing the position of the ISM gas, and the other one the halo of dark matter (see {\em e.g.}
\cite{Chardonnet:1996ca}). We take these slabs to have
radial extension $R_g$ and
heights $2 h_g$ (galactic disk) and  $2 h_h$ (halo) respectively. Furthermore, the flux of cosmic rays is supposed to be time independent over the relevant time scale, so that only a stationary solution of the diffusion equation is required.
There exist various models of anti-proton propagation, and they are all more or less in agreement with each others and with the  data, including PAMELA
\cite{Chardonnet:1996ca,Bottino:1998tw,Bergstrom:1999jc,Moskalenko:2001ya}. 

In the present article, to compute the flux of anti-protons at the top of the Earth  
atmosphere, we have used the propagation model implemented in the DarkSUSY package \cite{Gondolo:2004sc}. We have interfaced DarkSUSY with microMEGAs2.2 \cite{Belanger:2008sj, Belanger:2007zz}, a versatile package that allows to compute the dark matter abundance (and the relevant branching ratios) of any  WIMP dark matter candidate in models with a $Z_2$ parity.  

In the DarkSUSY code, the stationary anti-proton densities are taken to satisfy the transport
equation
\begin{equation}
\label{transport}
\frac{\partial n_{\bar{p}}}{\partial t}(E,\vec x)=\vec{\nabla} \cdot \left( D(R,\vec{x})~
\vec{\nabla} n_{\bar{p}} \right)  - \vec{\nabla} \cdot \left( \vec{u}(\vec{x})
~n_{\bar{p}} \right) -p(E,\vec{x}) ~n_{\bar p} + Q(E,\vec{x})=0\,.
\end{equation}
The description of propagation of anti-protons is accomplished by the first three terms. The first one describes diffusion in the galactic magnetic field. In each part of the Galaxy the
diffusion coefficient $D$ is assumed to be isotropic  
\begin{equation}
D(\vec{x}) = D(z) = D_g \theta(h_g -|z|) + D_h \theta(|z|-h_g),
\end{equation}
and to depend only on the magnetic rigidity parameter $R$ of the anti-protons (in GigaVolts or GV), 
\begin{equation}
D_l(R)=D^0_l\left( 1+\frac{R}{R_0} \right)^{0.6},
\end{equation}
where $l=g,h$ and $R_0  \sim 1$\,GV. The second term represents large scale convective motion, with a velocity field $\vec u(\vec x)$. This term is introduced to model the impact of the wind of cosmic rays, that blows away from the disk, on the motion of anti-protons. The third term, finally, models the loss of anti-protons due
to collisions with the ISM, mostly hydrogen, with
\begin{equation}
p(E,\vec{x})=n_H(\vec{x})~v(E)~\sigma(E)\,,
\end{equation}
where $\sigma$ is the inelastic cross section, and $n_H$ is
the number density of hydrogen in the galaxy, which is assumed to be  of the form
\begin{eqnarray} 
n_H(\vec{x}) = n_H(z) = n_{H_g} \theta(h_g -|z|) + n_{H_h} \theta(|z|-h_g)\,.
\end{eqnarray}
As for boundary conditions, it is assumed that the density of cosmic rays, and thus of anti-protons, is negligible at the boundary of the Galaxy, {\em i.e.}
$$
n_{\bar p}(R_g,z)= n_{\bar p}(r,h_h)=n_{\bar p}(r,-h_h)=0\,.$$

Finally, the last term on the r.h.s. in (\ref{transport}) is the source of anti-protons with energy $E$. Possible sources include secondary anti-protons produced by spallation and, possibly, dark matter annihilation or decay in the galactic halo. 
To discriminate between standard, astrophysical sources of anti-protons, and more exotic possibilities, like dark matter, it is clearly necessary to have a good handle on the expected background signal. 
For low energies, the transport 
models sketched above give a flux of anti-protons that is in good
agreement with observations, in particular if 
$p-He$ collisions and energy loss effects during propagation are taken into account \cite{Bergstrom:1999jc}.
However, the uncertainties on the various parameters (transport coefficients, nuclear processes,...) imply that
the background flux at high energies may not be uniquely determined from the low energy data\cite{Bringmann:2006im}. Nevertheless, for the sake of the argument, and like in most analysis of antimatter in cosmic rays, we  assumed that the background is known. More specifically, we adopt the flux used in the recent analysis of \cite{Bringmann:2006im}
\begin{equation}
\frac{d \phi}{d T_{\bar p}} = \frac{0.9
~t^{-0.9}}{14+30~t^{-1.85}+0.08 ~t^{2.3}} \quad[\mbox{GeV$^{-1}\cdot$m$^{-2}\cdot$
s$^{-1}\cdot$ sr$^{-1}$}]\,,
\label{eq:antiprotons}
\end{equation}
where $T_{\bar p} \equiv E_{\bar p} - m_{\bar p}$ and $t=T_{\bar p}/1$ GeV. 

The anti-proton data are generally presented in terms of the ratio of the anti-proton to proton fluxes, 
\begin{equation}
\label{backp}{\bar p\over p }\equiv {d\phi/dT_{\bar p} \over d\phi/dT_p}\,.\end{equation}
Thus we need the flux of protons. For the background, we use the spectrum \cite{Lionetto:2005jd} 
\begin{equation}
\frac{d \phi}{d T_{p}} = \frac{0.9~t^{-1}}{8+1.1~t^{-1.85}+0.8 ~t^{1.68}}\quad[\mbox{GeV$^{-1}\cdot$m$^{-2}\cdot$
s$^{-1}\cdot$ sr$^{-1}$}]\,.
\label{eq:protons}
\end{equation}
The ratio $\bar p/p$ observed is small, typically ${\cal O}(10^{-4})$, thus it is sufficient to only consider the background proton flux in the ratio (\ref{backp}).

\subsection{Positrons}
Like anti-protons, positrons in cosmic rays are supposed to be secondaries. Positrons may be created  in the decay of pions and kaons, which themselves are generated through
spallation. Like anti-protons, the motion of positrons in the galactic magnetic field is diffusive. They may also loose energy through
synchrotron radiation  in the interstellar magnetic field,  and through inverse Compton
scattering on diffuse starlight or on the cosmic microwave background. 

We follow here the classic treatment of \cite{Baltz:1998xv}.
The positron number density per unit energy, $dn_{e^+}/dE_{e^+}$,
is given by the stationary solutions of a transport equation 
\begin{equation}
\frac{\partial}{\partial t}\frac{dn_{e^+}}{dE_{e^+}}= \vec{\nabla}\cdot
\left[K(E_{e^+},\vec{x})\vec{\nabla}\frac{dn_{e^+}}{dE_{e^+}}\right]+
\frac{\partial}{\partial E_{e^+}}\left[b(E_{e^+},\vec{x})
\frac{dn_{e^+}}{dE_{e^+}}\right]+Q(E_{e^+},\vec{x})\,,
\label{eq:diffloss}
\end{equation}
where $K(E_{e^+},\vec{x})$ is the diffusion constant, $b(E_{e^+},\vec{x})$ is
the rate of energy loss, and $Q(E_{e^+},\vec{x})$ is the source
term (in cm$^{-3}$ s$^{-1}$). 
The relation between the flux and the density number is
\begin{equation}
\frac{d\Phi}{dE_{e^+}}= \frac{\beta c}{4 \pi} \frac{dn_{e^+}}{dE_{e^+}}.
\end{equation}
It is generally assumed that the diffusion constant is independent of position within the diffusion zone, and its energy dependence is taken to be given by 
\begin{equation}
K(E)=K_0 \left[C+ \left(\frac{E}{\mbox{1\,GeV}}\right)^{\alpha}\right],
\end{equation}
where $K_0=3\cdot10^{27}$cm$^{2}$s$^{-1}$, $C = 3^\alpha$ implements a plateau in the diffusion constant below $3$ GeV, and the spectral index is $\alpha=0.6$.
\cite{Baltz:1998xv}. The energy loss rate depends on the positron energy
through  
\begin{equation}
b(E)=\left(\frac{E}{\mbox{1\,GeV}}\right)^2 \frac{1}{\tau_E}\,,
\end{equation}
where $\tau_E = 10^{16}$s  is the characteristic time for energy loss. 
The diffusion zone is modelled in this case by a slab of thickness $2L$ with $L=3$ kpc. 
The diffusion equation is solved by considering a leaky box model, with free escape
boundary conditions, \textit{i.e.} the cosmic ray density is set to zero on the boundary of the slab.

\bigskip
Again, to confront the contribution of dark matter
annihilation to positrons in cosmic rays, we need a good handle on
the expected $e^+$ background.  An extensive discussion of the many
uncertainties on $e^+$ production by spallation and their
subsequent propagation in the interstellar medium may be found in
\cite{Delahaye:2008ua}. 
A further issue is that, to interpret the data, it is also necessary to know the electron flux. This is because the positron data are generally presented in term of  the  positron fraction 
$$
{e^+\over e^+ + e^-} \equiv { \left.\frac{d\Phi}{dE}\right\vert_{e^+}\over \left.\frac{d\Phi}{dE}\right\vert_{e^+}+\left.\frac{d\Phi}{dE}\right\vert_{e^-}}\,.$$
This ratio permits to factor out the, energy dependent, acceptance of the detectors (supposed to be the same for electrons and positrons), and, to some extent, the effect of solar modulation, to be discussed below. 

Unfortunately, the electron flux spectrum is poorly known. First, it is difficult to simulate the $e^-$ flux in cosmic rays from first principles,
simply because there are many astrophysical sources of
electrons in the Galaxy, starting with our Sun. Second, as of today, the experimental situation is not very clear, as one may appreciate from the compilation of existing data  presented in
\cite{Casadei:2004sb}, which show a wide spread in spectral index and even flux. Things will improve in the near future, with the advent of new data, including those on the spectra to be released by PAMELA. In the meantime, we have to confront the fact that the electron flux uncertainties have a non-negligible impact on interpreting the positron flux data, as discussed in  \cite{Delahaye:2008ua}. In the present article, were we focus on particle physics predictions, we adopt a common practice and take a fixed background electron and positron  fluxes. Concretely, we follow \cite{Baltz:1998xv}, and use the following  background flux of positrons and electrons, 
\begin{eqnarray}
\left. \frac{d\Phi}{dE}\right\vert_{e^-,\mbox{
\begin{tiny}prim.\end{tiny}}}&=&\frac{0.16 ~\epsilon^{-1.1}}{1+11 ~\epsilon^{0.9}+3.2
~\epsilon^{2.15}} \quad [\mbox{GeV$^{-1}\cdot$ cm$^{-2}\cdot$ s$^{-1}\cdot$
sr$^{-1}$}] \label{eq:em1bg}\\
\left. \frac{d\Phi}{dE}\right\vert_{e^-,\mbox{
\begin{tiny}sec.\end{tiny}}}&=&\frac{0.70 ~\epsilon^{0.7}}{1+110~\epsilon^{1.5}+600
~\epsilon^{2.9}+580~\epsilon^{4.2}} \quad [\mbox{GeV$^{-1}\cdot$ cm$^{-2}\cdot$
s$^{-1}\cdot$ sr$^{-1}$}] \label{eq:em2bg}\\
\left.\frac{d\Phi}{dE}\right\vert_{e^+,\mbox{
\begin{tiny}sec.\end{tiny}}}&=&\frac{4.5 ~\epsilon^{0.7}}{1+650 ~\epsilon^{2.3}+1500
~\epsilon^{4.2}} \quad [\mbox{GeV$^{-1}\cdot$ cm$^{-2}\cdot$ s$^{-1}\cdot$ sr$^{-1}$}] 
\label{eq:epbg}
\end{eqnarray}
where $\epsilon = (E/1$GeV$)$.  As for $\bar p$, we compute the flux of positrons from IDM dark matter annihilation  using DarkSUSY. 

\subsection{Solar modulation}

Before closing this section, we mention  one further complication.  Low energy ($E \lsim 1-10$ GeV) cosmic rays (both $e^+$ and $\bar p$) are strongly affected when they enter the Solar System, because they may loose energy
by interacting with the solar wind and the solar magnetic field. This is generically called solar modulation. The standard approach assumes that solar modulation is independent of the sign of the charge, and thus the same for electrons and positrons (for discussion see
\cite{Baltz:1998xv,Moskalenko:1997gh}). Using the,  so-called, Gleeson and Axford analytical force-field
approximation \cite{Gleeson:1967,Gleeson:1968}, the flux
at the Earth $d\Phi_{\oplus}/dE_{\oplus}$ can be deduced from the flux
at the heliosphere boundary, or interstellar (IS) flux $d\Phi_{IS}/dE_{IS}$, by
\begin{equation}
\frac{d\Phi_{\oplus}}{dE_{\oplus}}=\frac{p_{\oplus}^2}{p_{IS}^2} \frac{d\Phi_{IS}}{dE_{IS}},
\end{equation}
where the energy at the heliosphere boundary is given by
\begin{equation}
E_{IS}=E_{\oplus}+|Ze|\phi\,.
\end{equation}
Here $p_{\oplus}$ and $p_{IS}$ are the momenta at the Earth and at the
heliosphere boundary respectively, $e$ is the absolute value of the electron
charge, and $Z$ is the charge of the particle, in
unit of $e$. In this approximation, the solar
modulation is completely determined by the rigidity parameter $\phi$,
{\it i.e.} the particle momentum divided by its charge, expressed in volts, $[\phi]=V$. 
Since this solar modulation effect is the same for positive and negative charge particles, it does not affect the $e^+$ fraction and the $\bar p/p$ ratio. 

However, the recent data depart from this simple picture, and reveal some effects which may be attributed to a sign dependent solar modulation effect. In particular, below $10$ GeV, the  PAMELA experiment has measured a substantially smaller positron fraction than the previous experiments (essentially CAPRICE and HEAT  \cite{Boezio:1999hd,Barwick:1997ig,Duvernois:2001bb}). Similarly, over the years, BESS has measured variations in the $\bar p/p$ ratio. This sign dependent solar modulation effect is believed to be related to inversion of the solar magnetic field polarity, that take place with a periodicity of 22 years, in phase with maximum of solar activity, that occurs with 11 years periodicity (for details, see for instance \cite{Clem:2003br} and \cite{Sergio}). Still, for the time being, there seems to be no complete understanding of this extra solar modulation effect. This of course somehow limits the use we could make of low energy data (both on $e^+$ and $\bar p$), in particular to constraint models of dark matter. However we will try to argue otherwise.

\section{Dark Matter annihilation}

We now consider specifically annihilation, in the galactic halo, of an IDM dark matter candidate into Standard Model particles. The kind of annihilation products one may expect will depend on the mass range being considered. However, in the simplest version of the IDM, there is no direct annihilation into lepton-antilepton pairs\footnote{An exception is a light WIMP candidate, $M_{H_0} \sim $ GeV, which may have a large branching ratio into $\bar{\tau}\tau$ pairs, but this range is not relevant for explaining the PAMELA and ATIC features.}, and {\em a fortiori} into positrons.  Instead, positrons (and anti-protons) will results from decay of primary annihilation products, like $W^+W^-$ or $\bar{b}b$ pairs. This basic feature implies that we may not expect an IDM candidate to give a very good fit to both PAMELA and ATIC observations, which instead favour dark matter candidates with a large branching ratio (BR) into leptons (see {\em e.g.} \cite{Cirelli:2008pk,ArkaniHamed:2008qn}). 

The annihilation of an IDM dark matter particle into positrons or anti-protons enters the diffusion equations through the source term, 
\begin{equation}
Q(E,(\vec x)) =  \sum_{f} \langle\sigma_{\mbox{\begin{scriptsize}ann\end{scriptsize}}} v
\rangle_{tot} \cdot BR_f \cdot \frac{dN_f}{dE} \cdot\frac{\rho^2_{DM}(\vec{x})}{M_{DM}^2}\,,
\end{equation}
where $ \langle\sigma_{\mbox{\begin{scriptsize}ann\end{scriptsize}}} v
\rangle_{tot}$, $ BR_f $
and $\frac{dN_f}{dT}$ are respectively the total (average) 
annihilation cross section times the relative velocity $v$, the BR into  a state $f$ and the fragmentation function of the final state
$f$ into positrons/anti-protons.

\begin{figure}[!t]
\center
\begin{tabular}{cc} 
\includegraphics[width=0.45\textwidth]{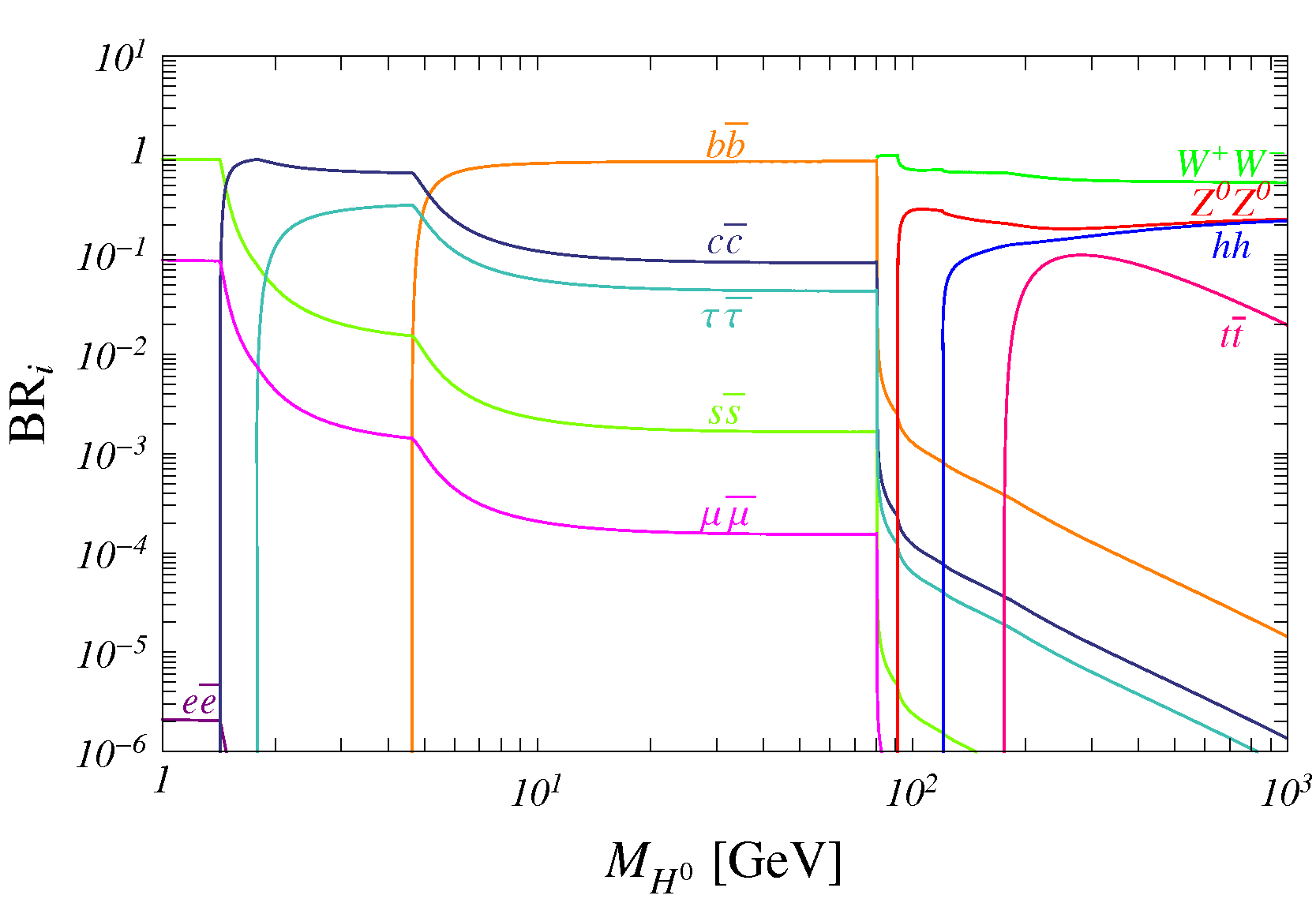}& 
\includegraphics[width=0.45\textwidth]{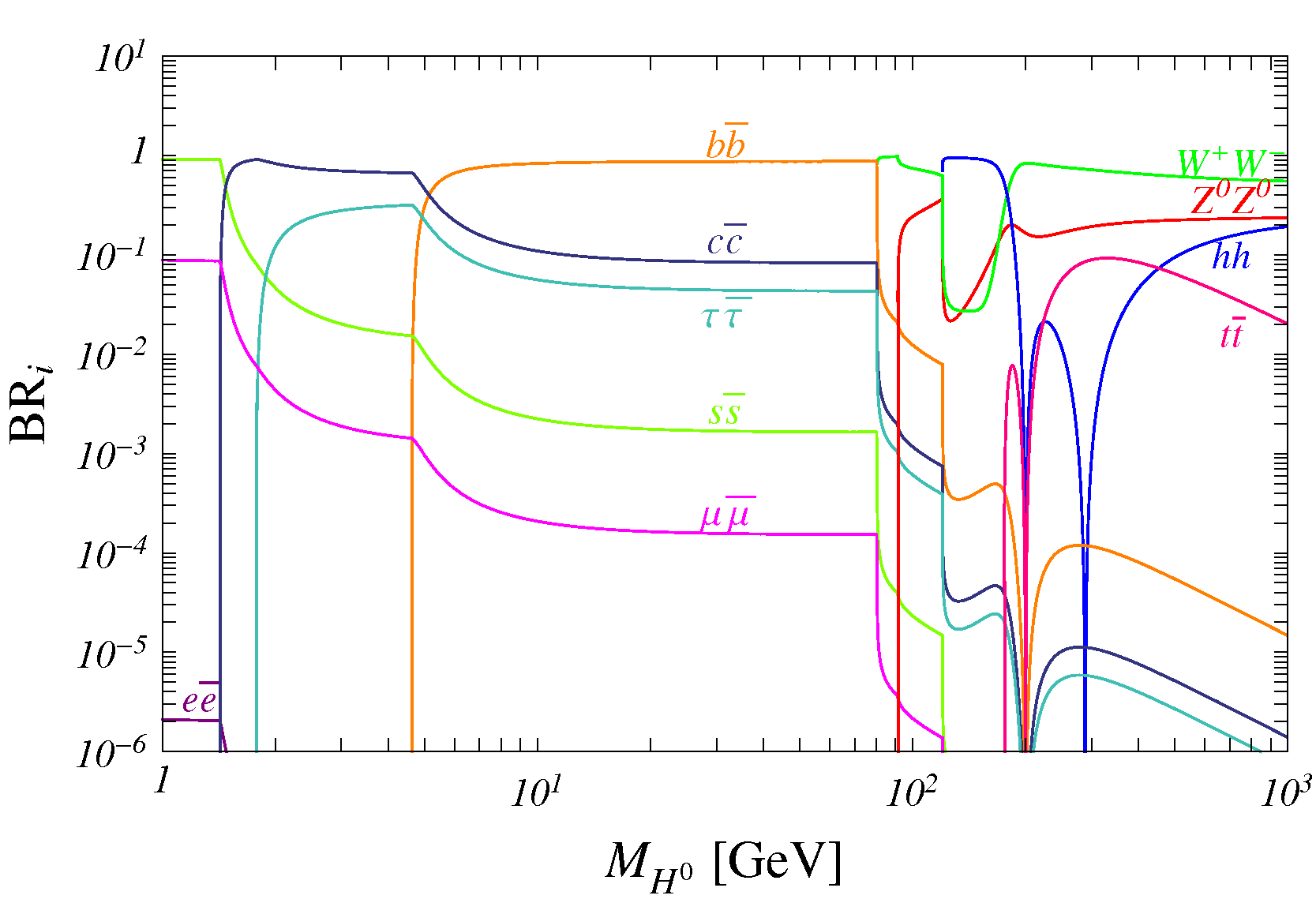}\\ 
\end{tabular} 
\caption{Branching ratios of the $H_0$ annihilation as a function of $M_{H_0}$. The Higgs mass $M_h$ is fixed at 120 GeV. \textit{Left panel}: $\mu_2=40$ GeV;
\textit{Right panel}: $\mu_2=200$ GeV.}.
\label{fig:branching} 
\end{figure}

We have implemented the IDM in the microMEGAs2.2 code, which gives the relic abundance of dark matter as a function of the model parameters. The code also gives the annihilation BR. Depending on the mass (and parameters), annihilation takes place dominantly either through the Higgs (low and middle mass ranges) or into $W^+W^-$ and $Z$ pairs (middle and high mass ranges). 
The BR are
shown in Fig.\ref{fig:branching}, for a Higgs mass $M_h=120$ GeV, and two values of the bare mass parameter, $\mu_2=40$ and $\mu_2=200$ GeV. From Fig.\ref{fig:branching}, we see that above 80 GeV the $H_0$ annihilates dominantly into weak gauge bosons. Once $M_{H_0} \geq M_{h}$, the annihilation into $h$ pairs opens, with a rate which depends on the $H_0-h$ coupling $\lambda_L$. 

The differences between the two panels of Fig. \ref{fig:branching} are perhaps worth being
explained. Below the threshold for weak gauge bosons production, the BR are the same in both figures. This is because, in this regime, the $H_0$ may only annihilate through the Higgs and the BR depend only on the Yukawa couplings of the fermions. Differences appear above the threshold. In particular, in the right panel, there is a dip in the fermion BR and two dips in the $hh$ BR. Both are related to  the change of the coupling $\lambda_L$, which is positive for $M_{H_0}  >
\mu_2$. In the right panel\footnote{In the left panel, $\lambda_L$ also vanishes, for $M_{H_0}=40$ GeV, but, below weak gauge bosons production, the BR do not depend on $\lambda_L$.}, the first dip corresponds to the vanishing of $\lambda_L$ at $M_{H_0}=\mu_2 = 200$ GeV. The second dip in $hh$ production is due to a destructive
interference, which may occur for $\lambda_L >0$ (see \cite{LopezHonorez:2006gr}). 

The rate of dark matter annihilation is proportional to the square of the number density
$n(\vec{r})=\rho(\vec{r})/M_{DM}$, which is not well known. Observations of rotation curves
suggest a rather cored profile \cite{Flores:1994gz,Kravtsov:1997dp}, with a flat behaviour at the
centre, whereas numerical simulations predict more cuspy
profiles in the innermost region of the galactic centre (see Kravtov et al. \cite{Kravtsov:1997dp},
Navarro-Frenk-White (NFW) \cite{Navarro:1995iw} and Moore et
al. \cite{Moore:1999nt}, as well as the recent Via Lactea and
Aquarius simulations \cite{Diemand:2008in,Springel:2008cc}). In the
present study, we focus on the  popular NFW distribution, which is a sort of benchmark in the community. The NFW profile is parameterised as follows
\begin{equation}
\rho(r) = \rho_0 ~\left(\frac{r}{r_0}\right)^{-\gamma} \left[
\frac{1+\left(r_0/a_0\right)^{\alpha}}
{1+\left(r/a_0\right)^{\alpha}}\right]^{\left( \frac{\beta-\gamma}{\alpha}\right)},
\end{equation}
where $r_0$ is the galacto-centric distance,  $\rho_0$ = 0.3 GeV/cm$^3$ is
the dark matter density in the solar neighbourhood, and (r$_0$ [kpc], a$_0$ [kpc],
$\alpha$, $\beta$, $\gamma$)= (8.0, 20, 1, 3, 1). 

On top of this smooth, averaged distribution, 
numerical simulations indicate that substructures or {\em clumps} of dark matter may survive in virialized systems
\cite{Silk:1992bh,Diemand:2006ik, Diemand:2007qr,Diemand:2008in,Springel:2008cc,Navarro:2008kc}, and it has
been suggested that such clumps may  enhanced  dark matter
annihilation rates. This effect is usually parameterised through a boost factor (BF)
\begin{equation}
\mbox{BF}= \frac{\int_V d^3x~\rho_{clumpy}^2}{\int_V d^3x~\rho_{smooth}^2}
\cong \frac{\int_V d^3x~\rho_{clumpy}^2}{\rho_{0}^2~V}\,,
\end{equation}
where the integration is over the volume which contributes to the
annihilation flux, typically a few kpc$^3$.  Recent developments tend to
disfavour the possibility of large enhancement of the fluxes
\cite{Lavalle:2006vb,Lavalle:1900wn,Lavalle:2008zb}, 
and, furthermore, it  has been shown that realistic boost factors have an energy dependence\cite{Lavalle:2006vb,Lavalle:1900wn}. Finally, it is expected to be different for positrons and anti-protons.
Despite these potential caveats, we follow the standard practice and
consider  boost factors  BF $={\cal O}(10-100)$ in this work. 

Another source of possible enhancement of the annihilation rates has recently received much attention. It has been shown that, in the presence of attractive long range interactions, the annihilation of non-relativistic dark matter particles may be enhanced in the limit of small relative velocities \cite{Hisano:2004ds}. This so-called Sommerfeld  {\em aka} Sakharov effect 
\cite{landau} may lead to spectacular enhancements, ${\cal O}(10-10^3)$, in particular near resonances. A general analysis of the Sommerfeld effect is quite complex \cite{Hisano:2004ds} but, a rule of thumb is that it is more relevant for heavy WIMPS, $M_{DM} \gsim $TeV. More precisely, the large mass IDM case  is very much analogous to the Minimal Dark Matter scenario, provided  $M_{DM} \gsim M_W/\alpha_W$, with $W$ boson exchange giving effectively  long range interactions with enhancements of ${\cal O}(10^2)$ \cite{Cirelli:2007xd}. For the low and middle mass range, however, the Sommerfeld effect is negligible. 

\section{The IDM {\em vs} the PAMELA, ATIC and HESS observations}
\label{sec:results}
The PAMELA satellite has published its first scientific results on $e^+$ and $\bar p$ in the Fall of 2008\cite{Adriani:2008zr}. The  primary goal of PAMELA is
the study of the antimatter component of moderate energy cosmic rays, with better resolution and, being a satellite and not a balloon experiment, with far better statistics than previous experiments.  

\begin{figure}
\begin{center}
\includegraphics[width=5.5cm]{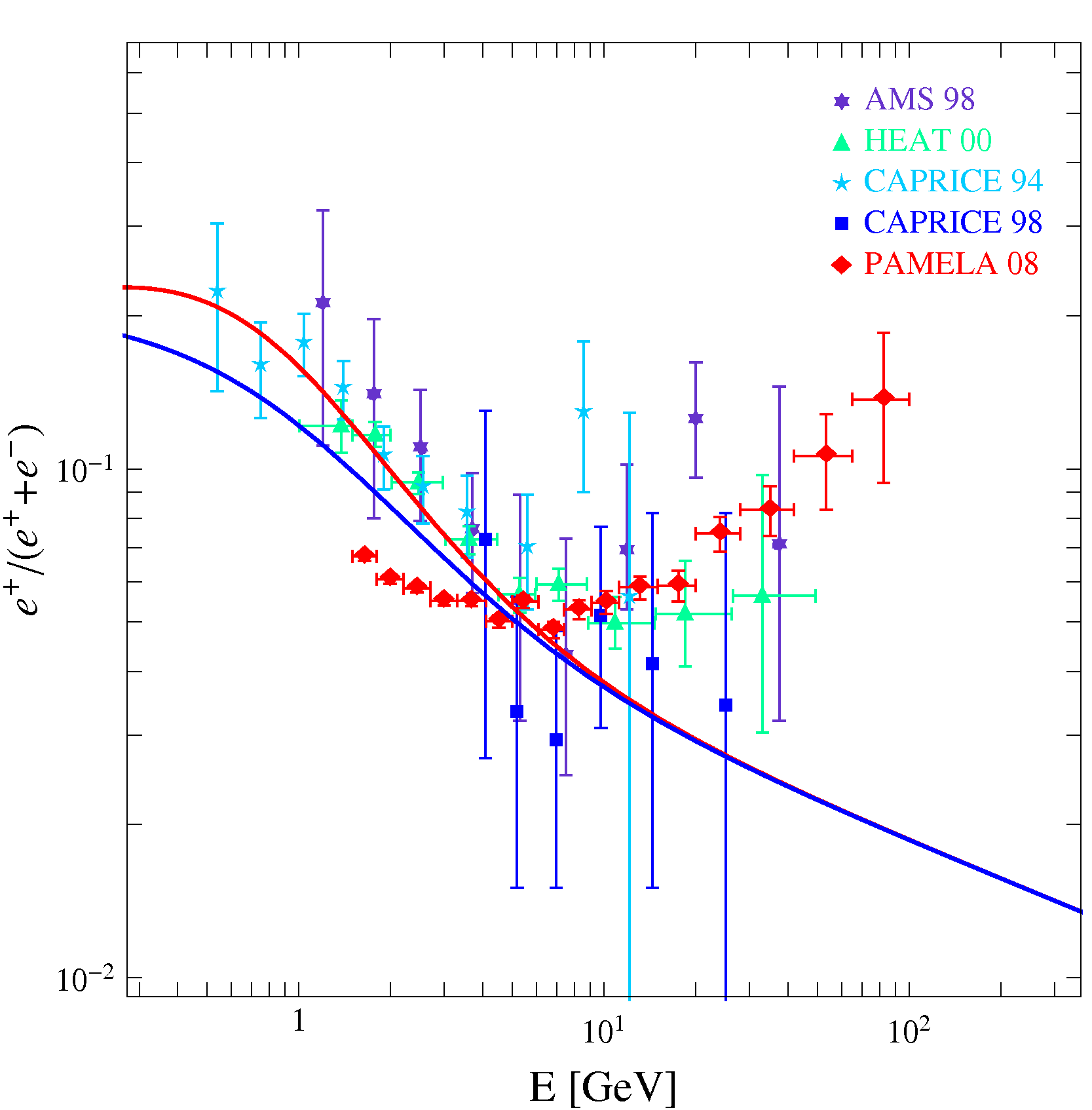}
\includegraphics[width=5.65cm]{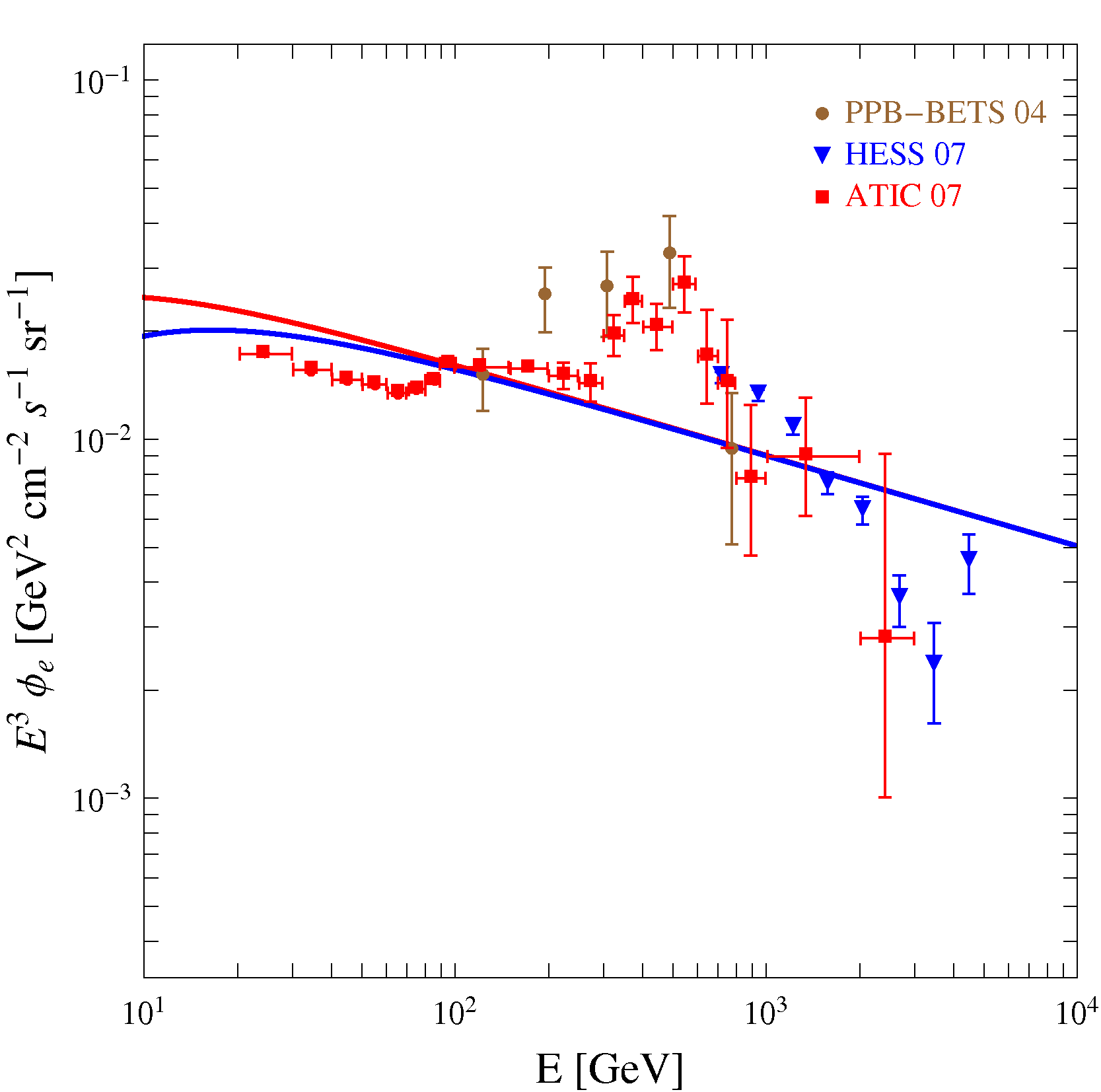}
\includegraphics[width=5.65cm]{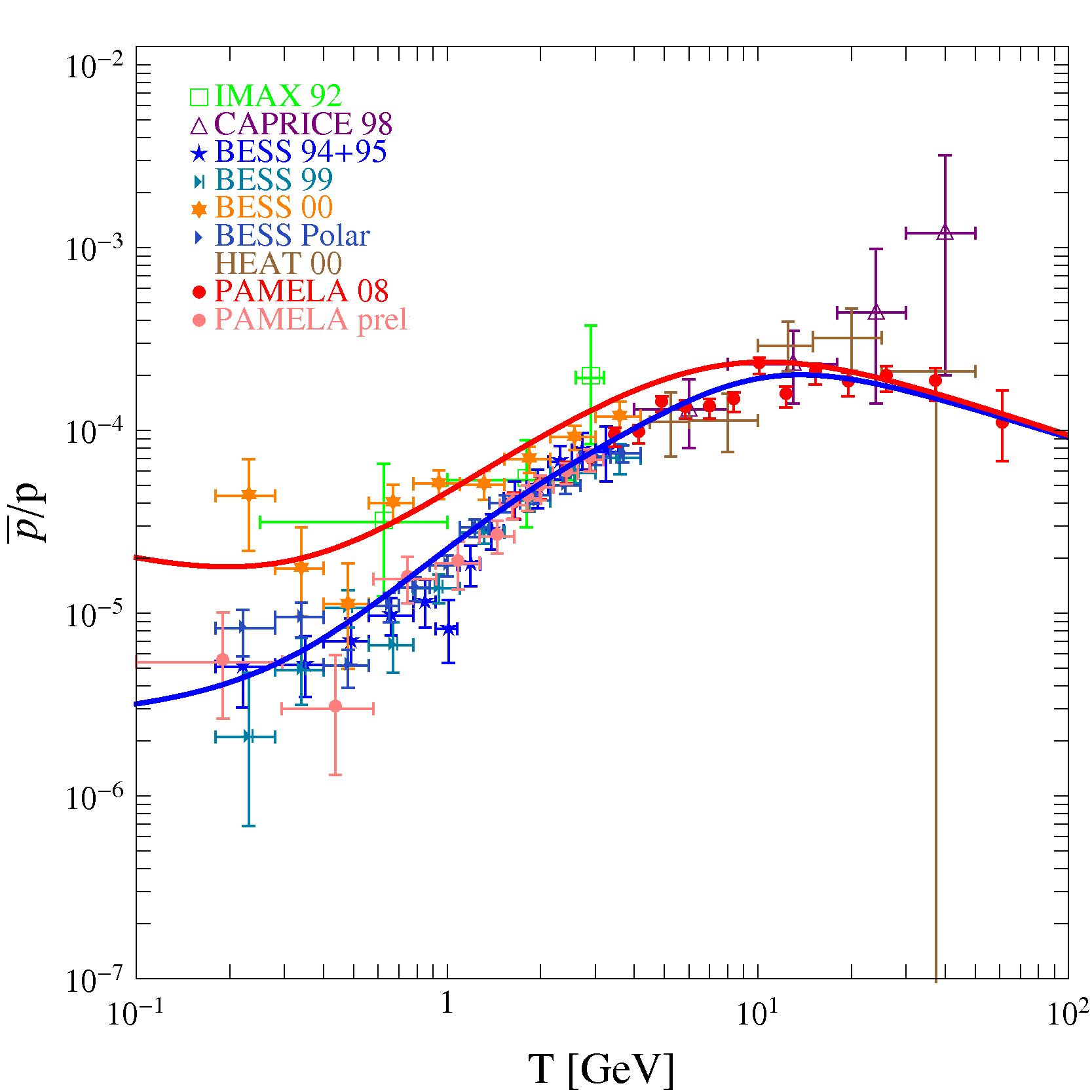}
\caption{\textit{Left panel}: Positron fraction measured by balloon
experiments and PAMELA (in red). \textit{Centre panel}: Total
electron and positron flux measured by balloon experiments, in particular ATIC, 
together with the HESS data. \textit{Right panel}: $\bar{p}/p$
fraction. In all cases the red (blue) lines represent the expected
background from Eq.~(13-15) and Eq.~(8-9) for leptons and baryons
respectively without (with) a solar modulation of $\phi = 500$\,MV. 
}
\label{fig:backgrounds}
\end{center}
\end{figure}

Concerning positrons, AMS 98 \cite{Aguilar:2007yf} and balloon experiments like CAPRICE 94 \cite{Boezio:2000ec}, CAPRICE 98 \cite{Boezio:1999hd} and HEAT 00 \cite{Beatty:2004cy},
have taken data between 300 MeV and 40 GeV. PAMELA has been designed to extend this range from
50 MeV to 270 GeV. The present data published so far cover the range between about 2 and 80 GeV (see Fig.\ref{fig:backgrounds}, left), corresponding to about $10^4$ positrons identified \cite{Adriani:2008zr}. 

The PAMELA collaboration has not yet released their observations of absolute flux of  $e^-$ and $e^+$, but the PPB-BETS, ATIC and HESS collaborations have measured the flux of electrons in the energy range 30\,GeV - 4\,TeV \cite{Torii:2008xu,Chang:2008ve,Collaboration:2008aaa}. Unlike PAMELA, these experiments can not distinguish electrons from positrons, hence in principle their data (Fig.\ref{fig:backgrounds}, middle) may also include positrons. The same figure also displays the recent HESS observations.

The anti-proton flux has been analysed by IMAX 92 \cite{Mitchell:1996bi}, BESS 95+97 \cite{Orito:1999re}, BESS 99 \cite{Asaoka:2001fv}, BESS 02 \cite{Shikaze:2006je}, BESS Polar \cite{Abe:2008sh}, CAPRICE 98 \cite{Boezio:2001ad} and HEAT 00
\cite{Beach:2001ub} experiments, from 120 MeV to 40 GeV. PAMELA has been designed to improve this measurement for energies from  80 MeV to 190 GeV. The present published data give the $\bar p/p$ ratio up to 100 GeV, corresponding to about $10^3$ $\bar p$ identified (Fig.\ref{fig:backgrounds}, right \cite{Adriani:2008zq, Sergio}).
Ultimately, PAMELA measurements will have a high statistics, with 10$^4$ (10$^5$) anti-proton (resp. positron) events after three years of operation. The data displayed in Fig.\ref{fig:backgrounds} are compared with the expected backgrounds, (\ref{eq:antiprotons}-\ref{eq:protons}) and (\ref{eq:em1bg}-\ref{eq:epbg}). For anti-protons, there is a good agreement between expectations and observations, but the positron and electron data show significant deviations. 

First, compared with previous experiments, the PAMELA data show a depletion of the $e^+$ fraction below $5-6$ GeV.
This discrepancy is not fully understood yet but, as discussed above, it is assumed to be due to a sign dependent solar modulation effect, related to the periodic inversion of the solar magnetic field. Naively, we would expect an anti-correlation in time between the positron and anti-proton fluxes ({\em i.e.} minima of positron fraction corresponding to maxima of $\bar p/p$ ratio) but, this is not what data on anti-proton show, Fig.~\ref{fig:backgrounds}. However, simulations of propagation of low energy cosmic rays indicate that light and heavy particles are affected differently by sign dependent solar modulation effects\cite{Clem:2003br}. This is manifest in the Fig.2 of \cite{Clem:2003br}, which shows that  both the $e^+$ fraction and $\bar p/p$ ratio should be low at the time of PAMELA data taking, which is actually quite in agreement with observations (see also the talk by S. Ricciarini \cite{Sergio}).

\begin{figure}
\center
\begin{tabular}{cc}
\includegraphics[width=0.44\textwidth]{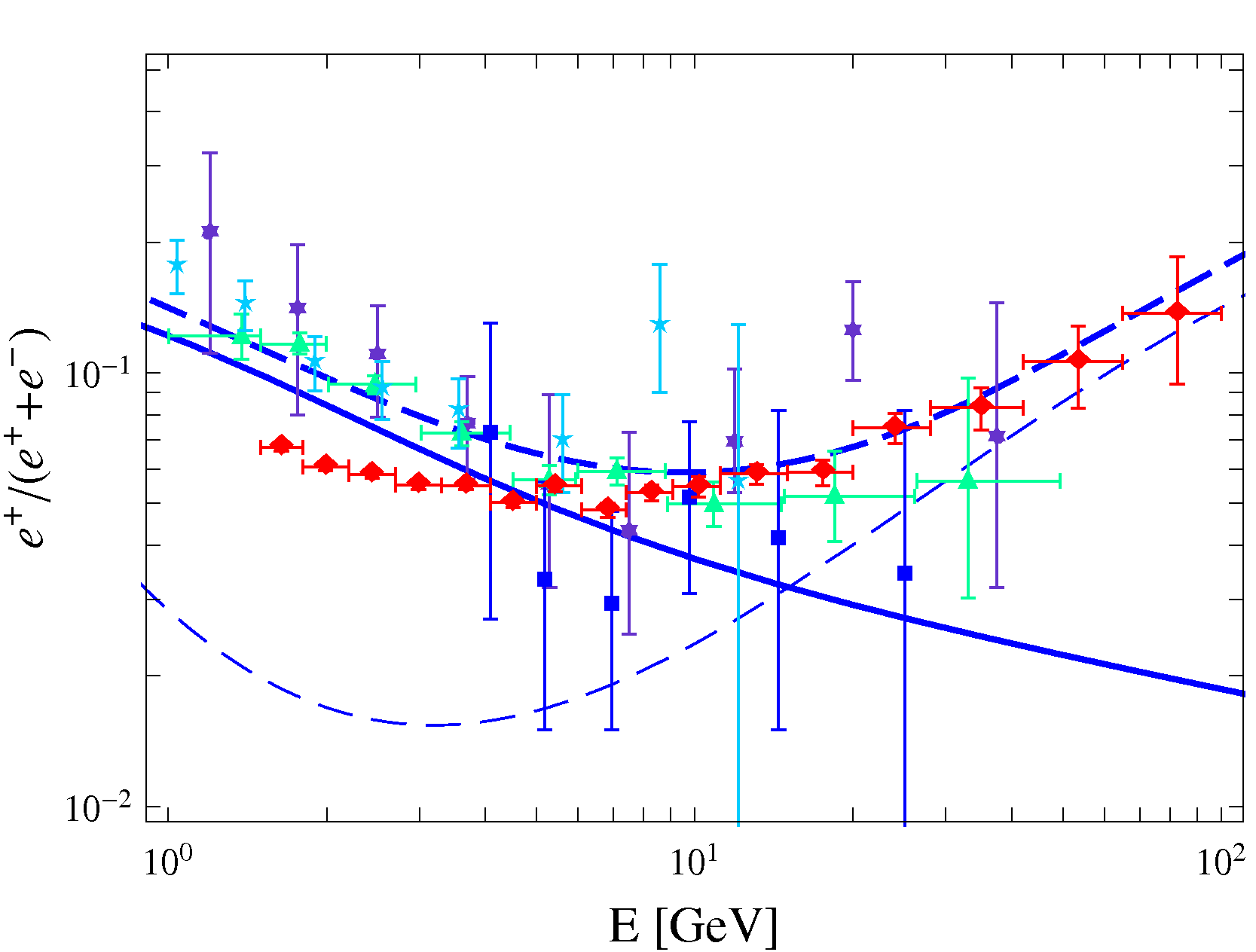}&
\includegraphics[width=0.45\textwidth]{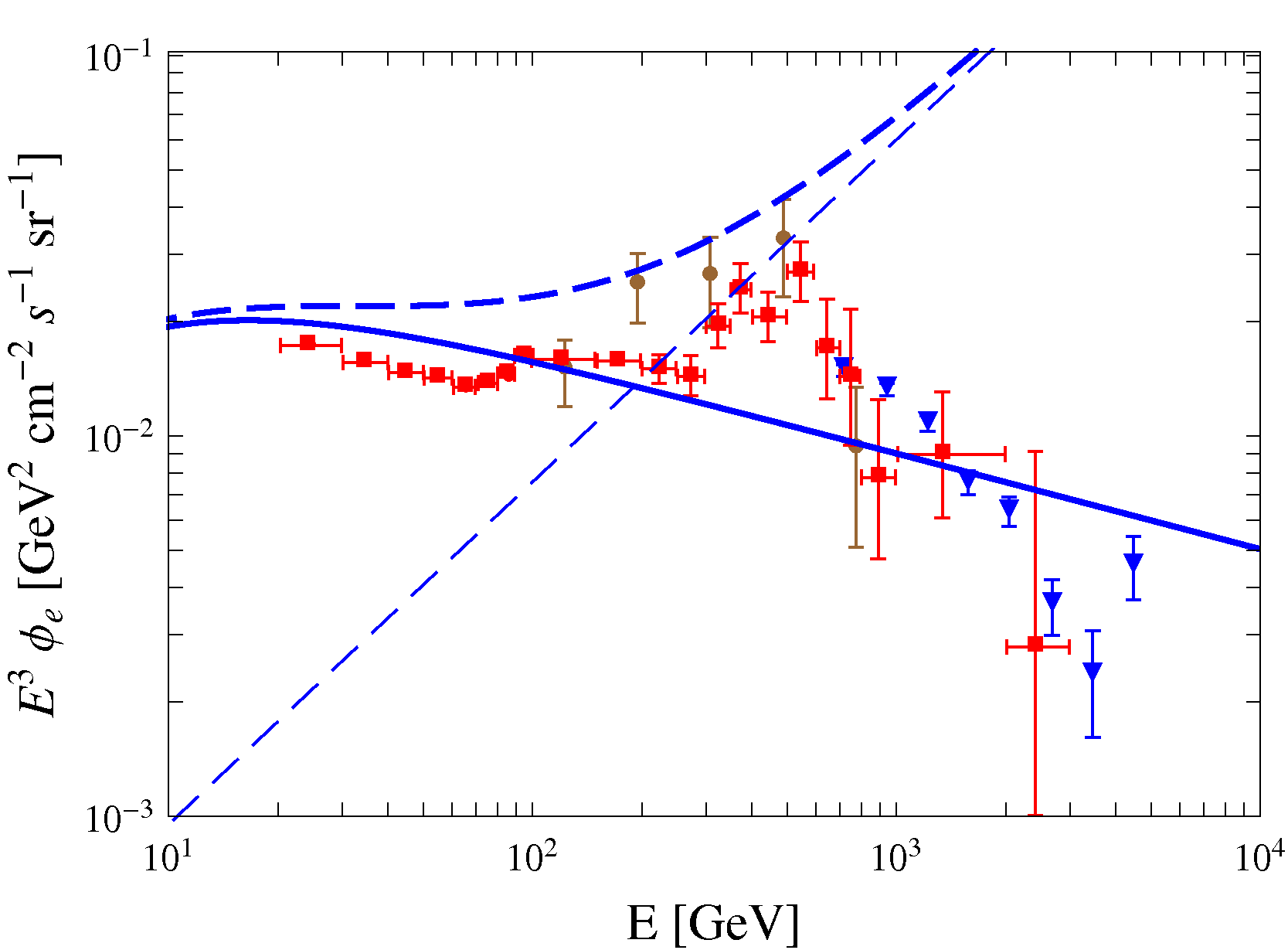}\\
\end{tabular}
\caption{\textit{Left panel} : The thin curve represents the positron fraction generated by a fiducial flux of positron $\Phi(E) = 6.5 \cdot10^5 \, E^{-2.1}$ GeV$^{-1}$ cm$^{-2}$ s$^{-1}$ sr$^{-1}$. \textit{Right panel} : Expected total electron plus positron flux in comparison with the ATIC measurements. In both panels, the \textit{thick plain} lines represent the backgrounds, whereas \textit{(thick) dashed} lines represents the signals (+ bgd).
}
\label{fig:extrapolation:epF}
\end{figure}

Second, the PAMELA data also show a steadily increase of the $e^+$ fraction  above $\sim 10$ GeV.  The standard expectation, based on simulations of $e^+$ production through spallation and propagation in the ISM, is that the $e^+$ fraction should decrease instead of increasing with energy. This is manifest in the expected flux of Eq.(\ref{eq:em1bg}-\ref{eq:epbg}) which give 
$$
\left.{e^+\over e^+ + e^-}\right\vert_{\mbox{\rm bckg}} \propto E^{-0.15}\,.
$$
The data (focusing on the three high energy points, see Fig. (\ref{fig:extrapolation:epF}) gives
$$
\left.{e^+\over e^+ + e^-}\right\vert_{\mbox{\rm PAMELA}} \propto E^{0.55}\,.
$$
This positron excess is the cause of the recent excitement around the PAMELA
data, a possible explanation being dark matter particles
annihilating in the vicinity of the solar system. Furthermore it is
tempting to correlate the PAMELA excess to the electrons+positrons excess observed by ATIC at energies in the $300-800$ GeV
range. This excess is also consistent with the HESS data.

Actually, if we fit the PAMELA data by adding to the background an extra positron flux, with a simple power law spectrum, and an equivalent electron flux, the excess in PAMELA and ATIC data may be put in correspondence. This is suggested in Fig.\ref{fig:extrapolation:epF}, were we used $\Phi_{e^-,e^+} \propto E^{- 2.1}$. Our flux somewhat overshoots the ATIC data, but this could be ameliorated by lowering the expected background (plain blue curve in the right panel of Fig.\ref{fig:extrapolation:epF}), something we were however reluctant to do. Obviously, the flux  should also be cut-off at an energy around the ATIC peak. Altogether, such a spectrum may result from  a source  with an energy injection around the ATIC excess. These basic features are somehow consistent with  the astrophysics explanation, which posits that the electron and positron excesses are due to cosmic rays produced by nearby pulsars \cite{Aharonian:1995he,Hooper:2008kg,Yuksel:2008rf,Profumo:2008ms,Shaviv:2009bu}. Explaining both PAMELA and ATIC data from dark matter is in general more challenging but, 
as discussed extensively in the recent literature, a very simple, albeit {\em ad hoc}, culprit would be a heavy dark matter particle that annihilates or decays (in the case of a long-lived dark matter candidate) dominantly in lepton pairs. 

In the sequel of this work we investigate  annihilation of an IDM dark matter into antimatter, and confront the model to the data.

\subsection{Low mass range}
In \cite{Andreas:2008xy} it has been shown that a light IDM dark matter candidate ({\em i.e.} a light WIMP) may be consistent with both the WMAP and the DAMA/LIBRA results. DAMA/LIBRA is a direct detection experiment that has reported evidence for an annual modulation of the nuclear recoils in their detector \cite{Bernabei:2008yi}. Taking into account the null results of all the other DM detection experiments (and the channelling effect on the threshold energy in DAMA), the DAMA/LIBRA signal may be due to elastic scattering of a dark matter from the galactic halo, with a dark matter particle mass in the range (see \cite{Petriello:2008jj}) \footnote{The best fit to DAMA/LIBRA data gives a somewhat larger mass
$M_{DM} \sim 10-12$ GeV, but these values are excluded by the CDMS and
Xenon data \cite{Fairbairn:2008gz,Chang:2008xa}.}
$$
3 \ \mbox{\rm GeV} \lsim M_{DM} \lsim 8 \ \mbox{\rm GeV}\,.
$$
\begin{figure}[!t]
\begin{center}
\includegraphics[width=5.5cm]{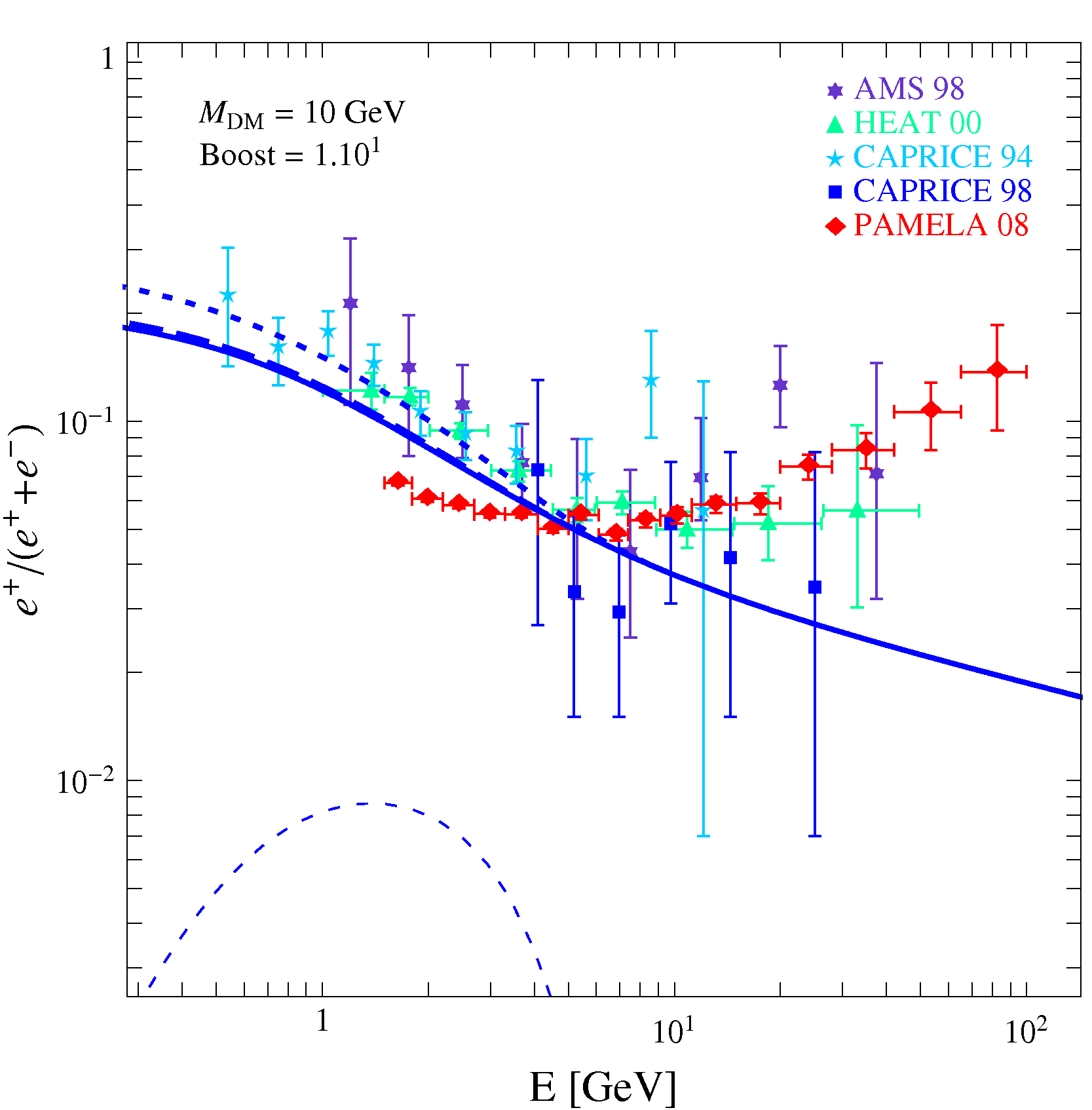}
\includegraphics[width=5.6cm]{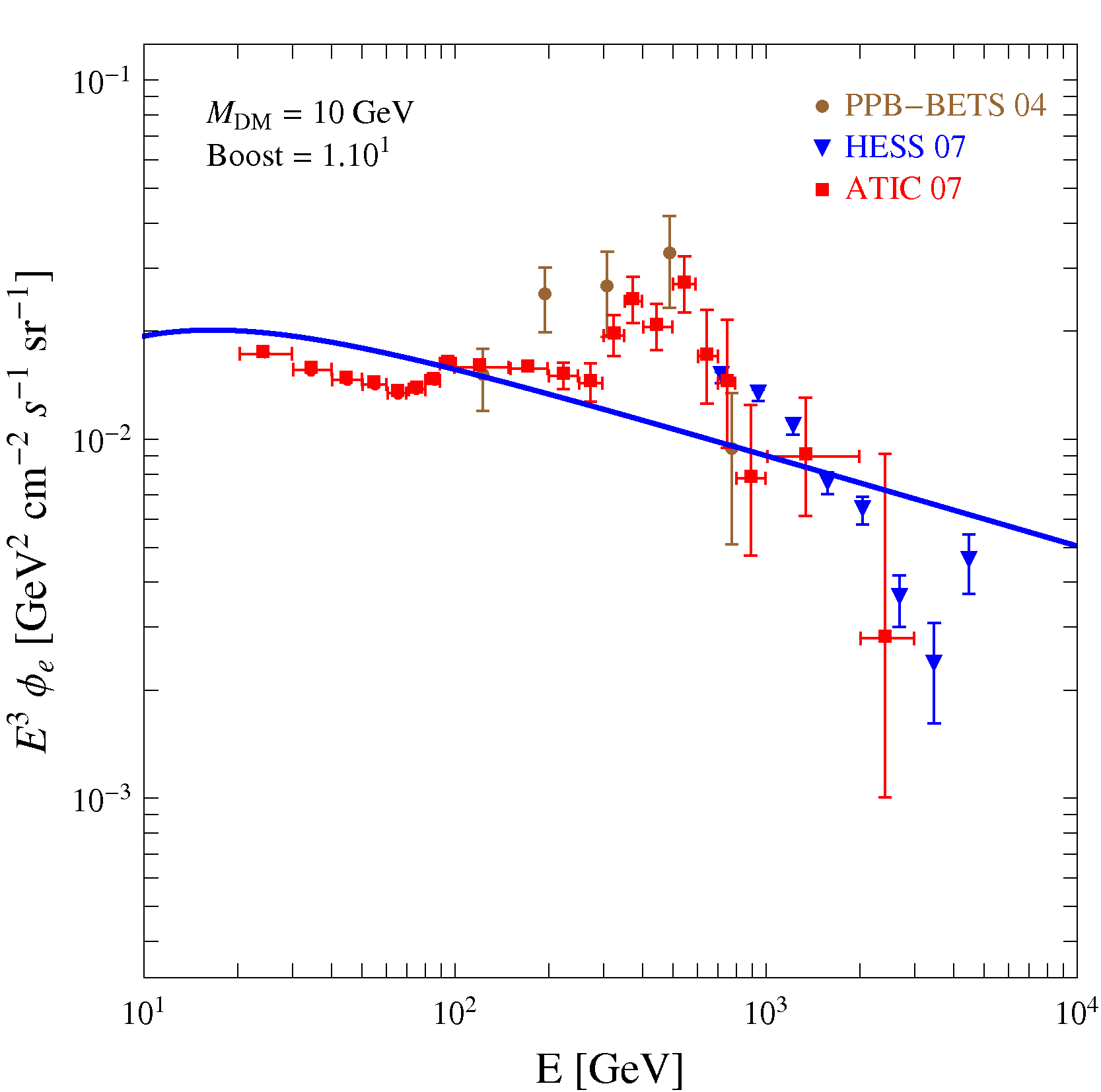}
\includegraphics[width=5.45cm]{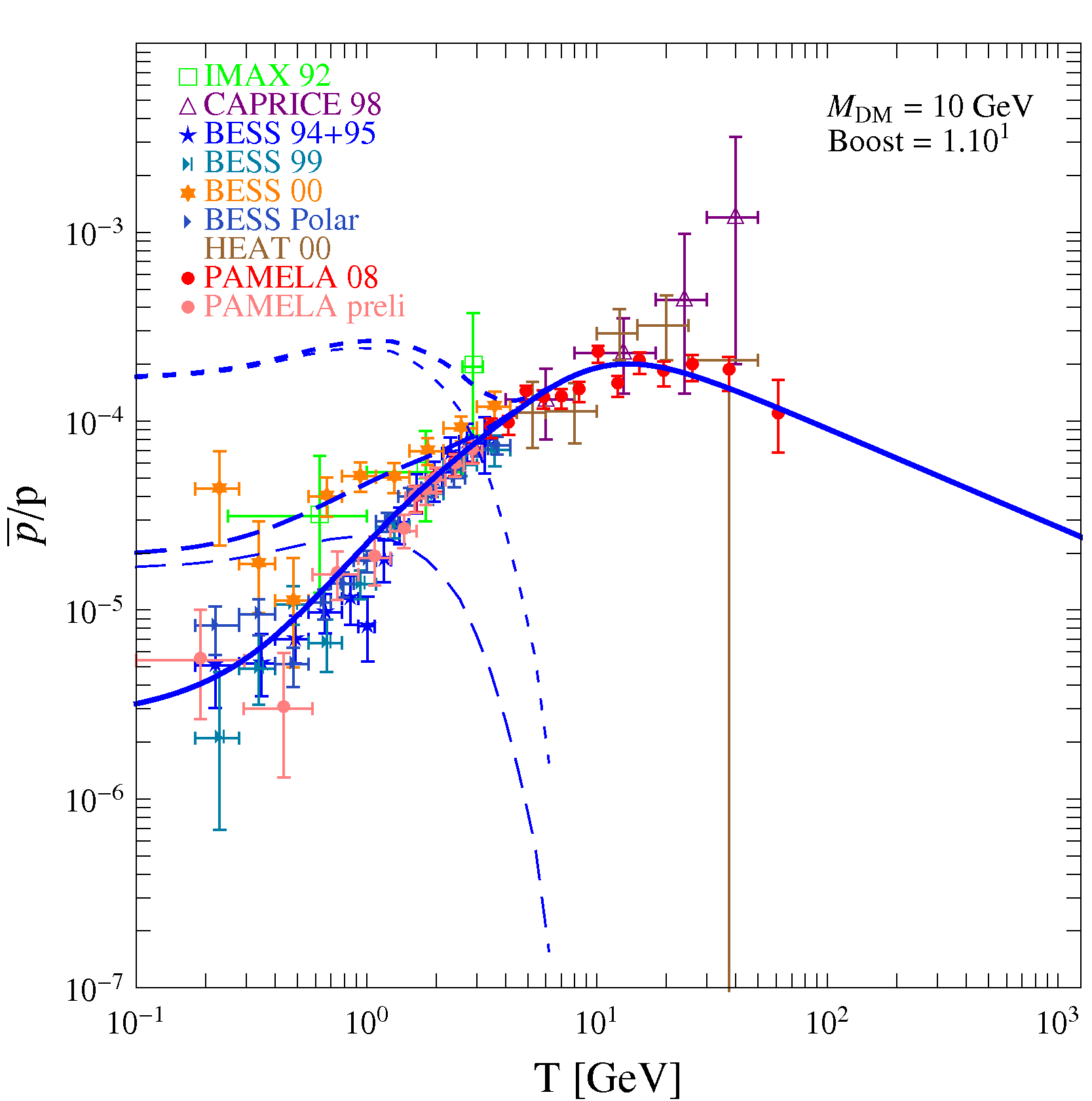}
\caption{A Low Mass candidate ($M_{DM}=10$\,GeV).  \textit{Thick Plain}: Background, \textit{(Thick) Dashed}: Signal (+ BckGrd) and \textit{(Thick) Dotted}: Signal with boost factor = 10 (+ BckGrd). A solar modulation of $\phi = 500$~MV has been applied.}
\label{fig:lowmass}
\end{center}
\end{figure}
\begin{figure}[!t]
\begin{center}
\includegraphics[width=5.6cm]{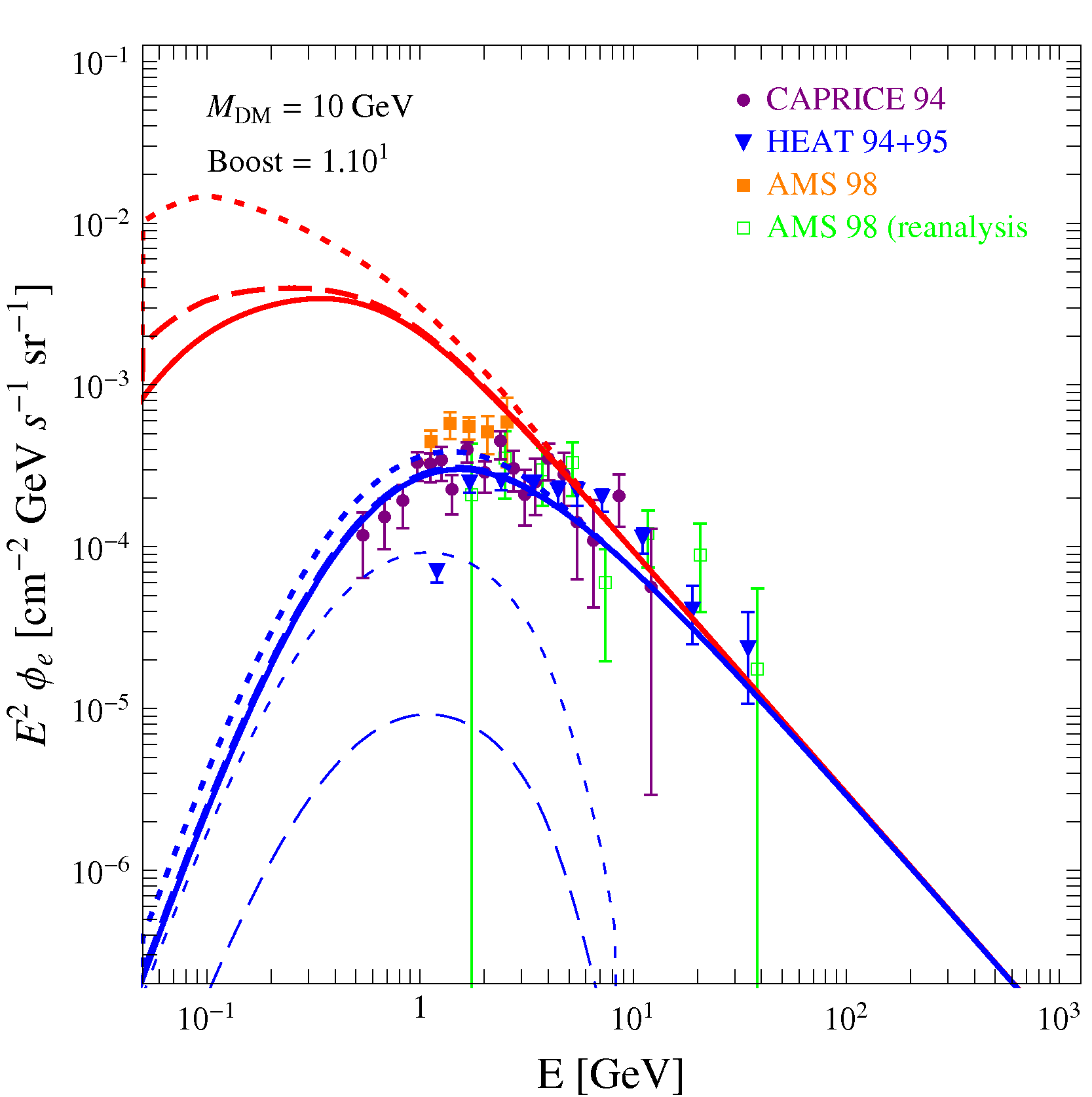}
\includegraphics[width=5.68cm]{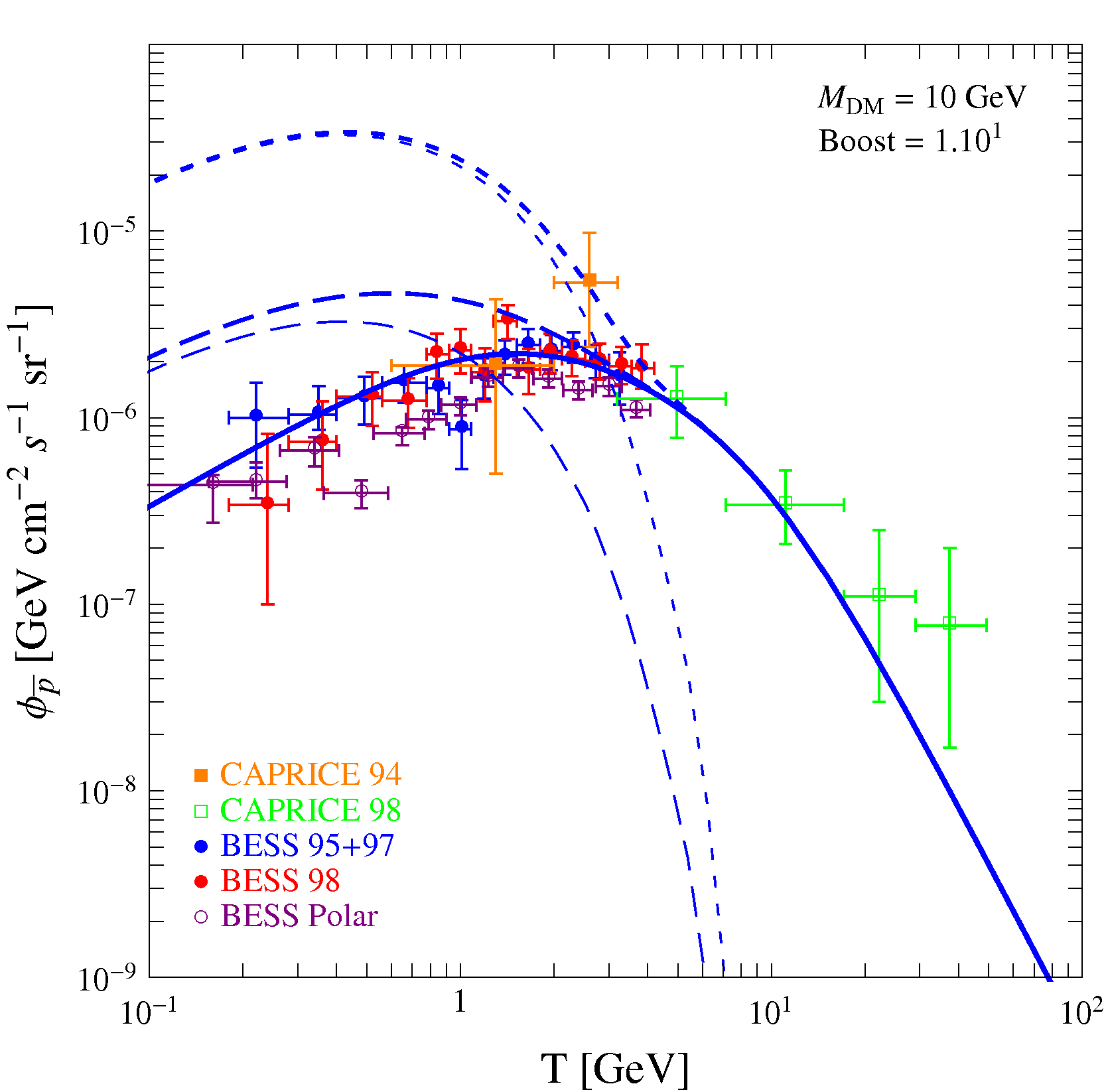}
\caption{\textit{Left panel} (\textit{Right panel}) : Flux of positrons (anti-protons) for $M_{DM}=10$\,GeV. \textit{Thick Plain}: Background, \textit{(Thick) Dashed}: Signal (+ BckGrd) and \textit{(Thick) Dotted}: Signal with boost factor = 10 (+ BckGrd). The red (blue) curve corresponds to signals without (with) solar modulation ($\phi = 500$~MV).}
\label{fig:lowmass2}
\end{center}
\end{figure}

For these
masses, both annihilation of the $H_0$ and its scattering with a nucleus
take place through the Higgs particle,  with  cross sections scaling
like $\sigma \propto \lambda_L^2/m_h^4$. Hence, for a fixed DM mass, imposing
WMAP abundance fixes all the  relevant parameters of the model (modulo
some residual uncertainty in the Higgs-nucleus couplings). That
agreement with DAMA/LIBRA may be reached without further tuning is thus remarkable. This prediction is however not specific to the IDM. As emphasised in \cite{Andreas:2008xy}, any model with a scalar DM candidate coupled through the Higgs portal would give a similar fit to the data. 

In our discussion of positron and anti-proton signatures we consider as
a benchmark, a light WIMP scalar candidate with mass $M_{DM} = 10$ GeV and $\sigma_{SI}
\approx 3\cdot 10^{-41} \, {\rm cm}^2$.  This candidate is somewhat within the boundary between the ranges allowed by DAMA/LIBRA and that excluded by all other experiments. However, the version of DarkSUSY code that we have adapted does not give the flux of positrons and anti-protons for a dark matter candidate lighter than 10 GeV. Computing the flux for the whole DAMA/LIBRA range would require to update the Pythia tables of DarkSUSY toward lower masses, but we have not done so. Still, we do not expect much difference between  the flux produced by 10 GeV and 8 GeV dark matter candidates. Substantially lighter candidates, say with $M_{DM} \sim 4 $ GeV, could give larger flux (albeit at lower energies), but these candidates are less favoured by the DAMA/LIBRA data. So, at this exploratory level, we limit ourself to a candidate with $M_{DM} = 10$ GeV. 

A cosmic abundance consistent with WMAP is reached for $\lambda_L
\approx -0.2$ for $M_h=120$ GeV and  $M_{DM} = 10$ GeV. Consequently, the flux
of positrons and anti-protons is  fixed,  modulo the usual uncertainties
regarding the  distribution of dark matter in the halo or in 
the cosmic ray  propagation parameters. The results are given in Fig.\ref{fig:lowmass}. In all three figures, 
the dashed line corresponds to a flux computed for a standard NFW distribution, while the dotted line correspond to a signal boosted by a factor of 10. Given the low mass of the $H_0$, the signal is within the range of energies were both the positron and anti-proton fluxes are subject to solar modulation. This complicates the comparison between  theoretical predictions and data. 

For both positrons and anti-protons, the contamination from dark matter annihilation is consequent only if we consider boosting the signal. Here we take BF = 10. Also the effect of the boost is  more dramatic for anti-protons than for positrons. This is essentially  because the expected background signal is much smaller for $\bar p/p$ than in the case of the positron fraction.  
Focusing on positrons, the excess from dark matter annihilation may be somewhat attenuated (hidden) by playing with the rigidity parameter $\phi$ (we have used $\phi=500$ MeV in the plots), but doing so would also affect all low energy cosmic rays fluxes (including heavier nuclei) so we have refrained doing so. 
This still leaves open the possibility of hiding the dark matter signal with a sign dependent solar modulation effect (provided the boost factor is not too large). It remains to be seen whether this can be done in a consistent way for all cosmic ray constituents, which differ in mass or charge. Moreover,  dark matter annihilates into fermion-antifermion pairs, and so an excess would also be manifest in absolute fluxes. These are shown in Fig.\ref{fig:lowmass2} for positrons and anti-protons, together with existing data  AMS 98 \cite{Aguilar:2007yf}, CAPRICE 94 \cite{Boezio:2000ec}, CAPRICE 98 \cite{Boezio:1997ec}, BESS 95+97 \cite{Bergstrom:2008gr}, BESS 98 \cite{Maeno:2000qx} and BESS Polar \cite{Abe:2008sh}. In the absence of a systematic analysis of (sign dependent) solar modulation in the light of recent data, for the time being it is probably fair to conclude that little may be said from positrons data alone. However, the current data on anti-protons  allow to exclude  BF larger than one. In the future, we expect constraints from positrons to remain much milder than those from anti-protons, because the DM contribution is typically smaller than the overall  effect of solar modulation. 
We should emphasised that BF = 10  is already a large boost factor, and {\em a fortiori} so
for anti-protons. Indeed, since anti-protons may travel much greater distances than positrons without
losing energy, the anti-proton flux is expected to be
less subject than the positrons to  clumps in the dark
matter distribution \cite{Lavalle:2008zb}. This means that the astrophysical BF is expected to be smaller for anti-protons than for positrons \cite{Lavalle:1900wn}.

To conclude the discussion on a the light IDM dark matter candidate, we briefly comment on the production of anti-deuteron. 
The production of anti-deuteron by spallation is typically small and is predicted to fall off for kinetic energies below 1 GeV per nucleon. Given its large abundance, a light WIMP may give a substantial contribution to the flux of anti-deuteron at low energies (see for instance  \cite{Donato:1999gy}). For the IDM candidate with $M_{DM}=10$\,GeV, we obtain, using the DarkSUSY routines, an anti-deuteron flux at $T_{\bar{D}} = 0.25$\,GeV/n of $9\cdot10^{-7}$ (GeV/n s sr m$^2$)$^{-1}$ (for BF = 1), which is below the upper limit of $1.9\cdot 10^{-4}$ (GeV/n s sr m$^2$)$^{-1}$ set by the BESS experiment \cite{Fuke:2005it}, but above the expected acceptance of the future AMS-02 and GAPS experiments \cite{Giovacchini:2007,Fuke:2008zz}, which are $4.5\cdot 10^{-7}$ (GeV/n s sr m$^2$)$^{-1}$ and $1.5\cdot 10^{-7}$ (GeV/n s sr m$^2$)$^{-1}$ respectively. Anti-deuteron data might turn out to give the strongest constraint on a light IDM dark matter candidate.

\subsection{Middle mass range}

\begin{figure}
\begin{center}
\includegraphics[width=5.5cm]{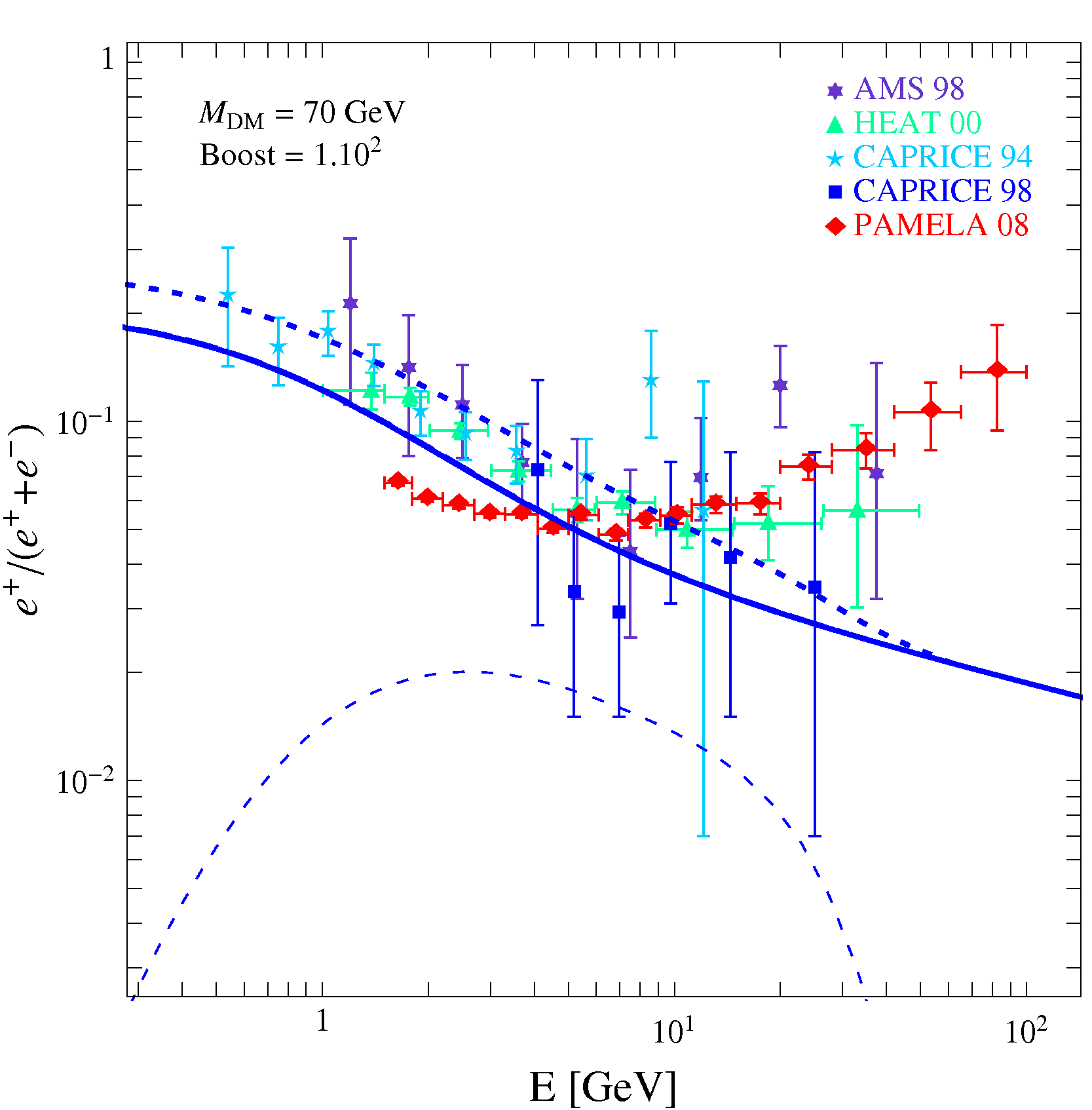}
\includegraphics[width=5.6cm]{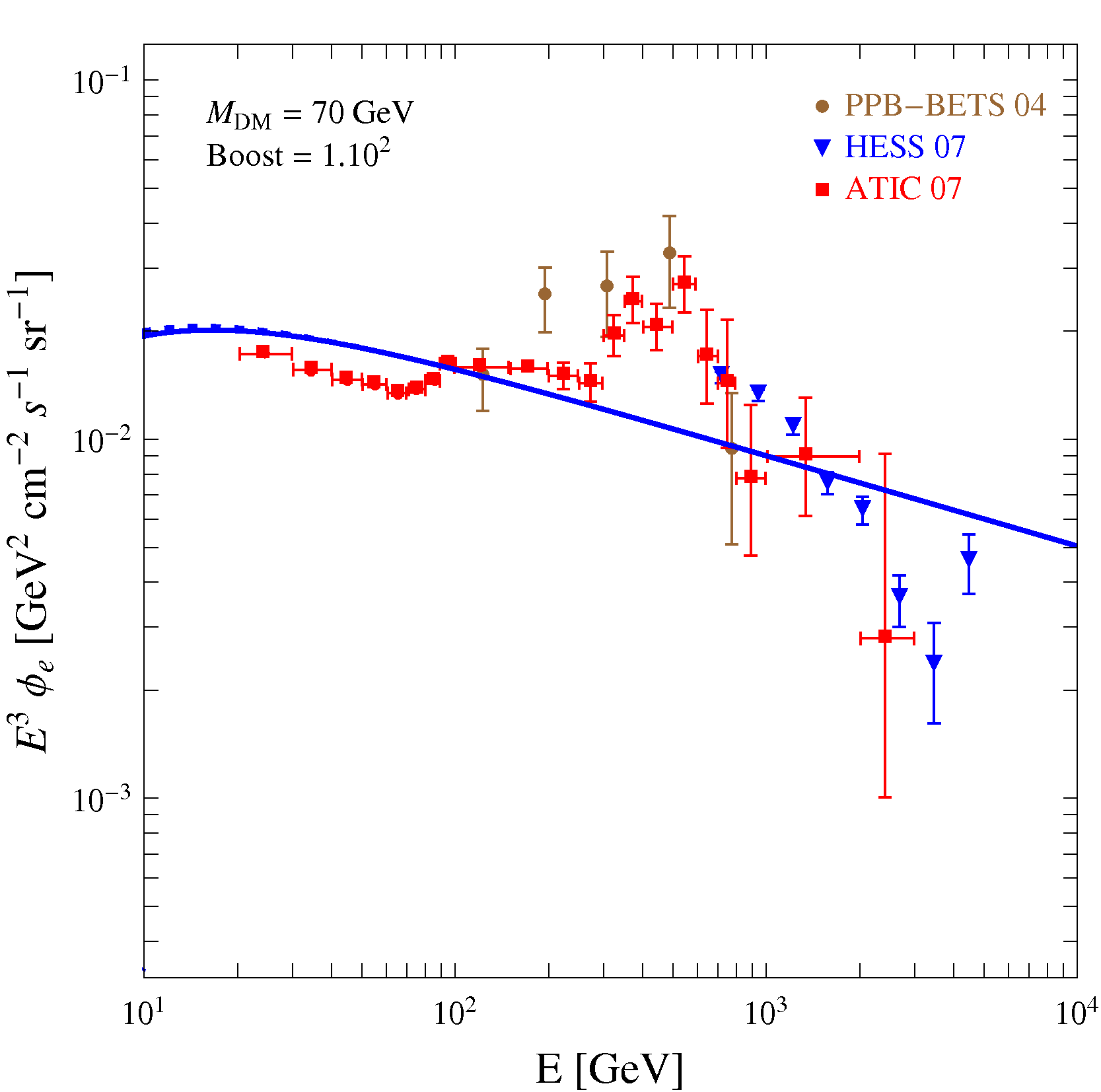}
\includegraphics[width=5.45cm]{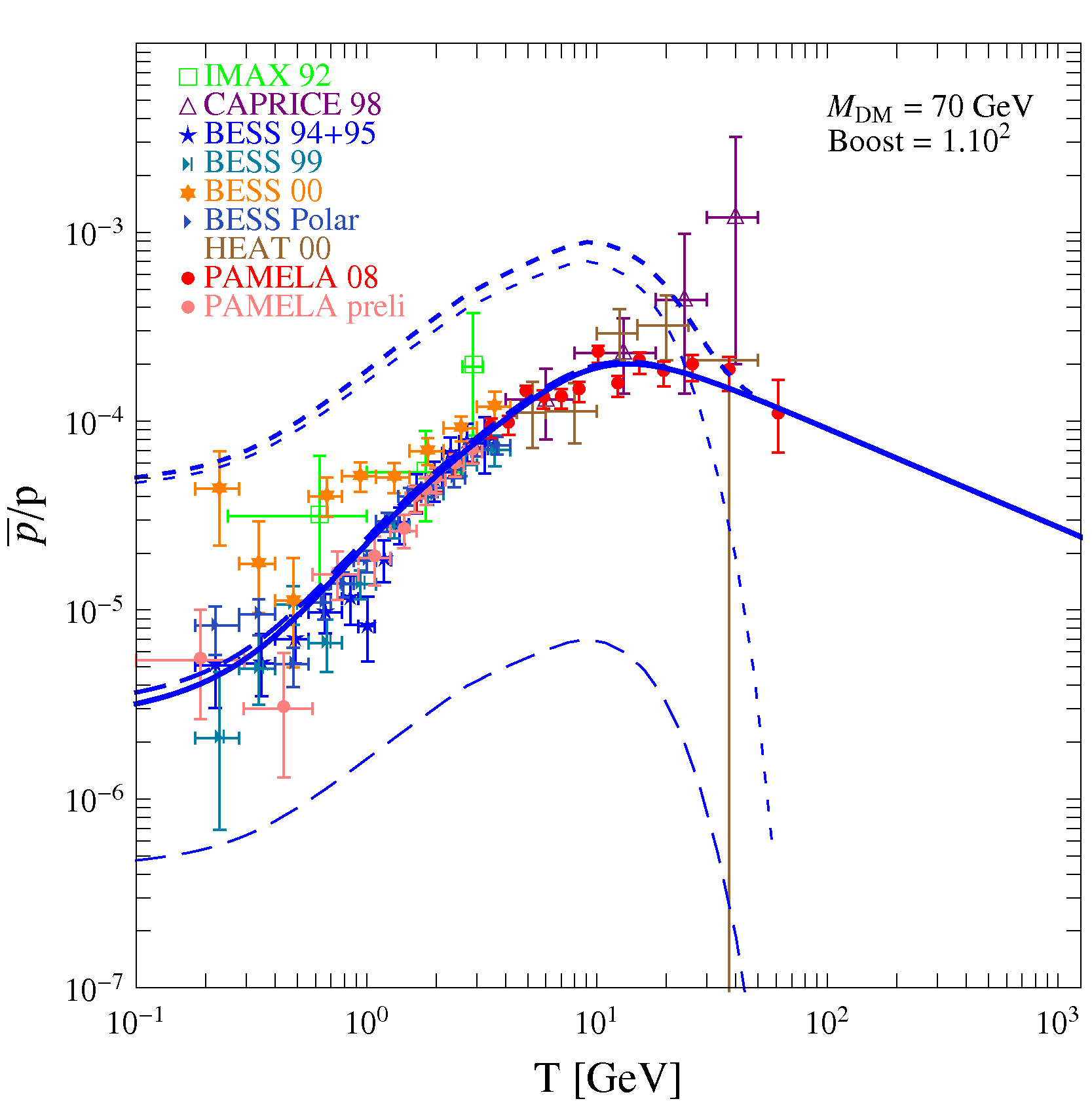}
\caption{Same as Fig.~\ref{fig:lowmass} but for a Middle Mass candidate ($M_{DM}=70$\,GeV, BF = 10$^2$). }
\label{fig:middlemass}
\end{center}
\end{figure}

In this subsection, we consider an IDM dark matter candidate in the mass range 
$$
40 \mbox{\rm\, GeV} \lsim M_{DM} \lsim 80 \mbox{\rm\, GeV}\,.
$$
In this range, both  annihilation through the Higgs and co-annihilation through the $Z$ are important to determine the relic abundance \cite{Barbieri:2006dq,LopezHonorez:2006gr}.
An illustration is given in  Fig.~\ref{fig:middlemass},  where the fluxes and fractions from the annihilation of a  70\,GeV candidate, which  has a cosmic abundance that is in agreement with WMAP.  

Since annihilations produce dominantly $\bar{b}b$ pairs, the positron
spectrum emerging is very soft. Even if one considers a large
boost factor to enhance the signal (BF =
10$^2$ in the figure), the spectrum does
not exhibit the steep increase in the positron fraction observed by PAMELA. There is a slight excess of positrons below 10 GeV, but it
is consistent with observations. Moreover this excess is within the
energy range where solar modulation is effective, and may thus be easily hidden.

Like in the case of a light WIMP, the excess of $\bar p$ is more significant than in positrons. 
Although the signal may also be mimicked by solar modulation for moderate boost factors, large boost factors  (say BF $\gsim10^2$) may be clearly excluded. Finally, there is a small excess around 10 GeV that perhaps could be probed with better statistics and better understanding of solar modulation effects.

\begin{figure}
\begin{center}
\includegraphics[width=5.5cm]{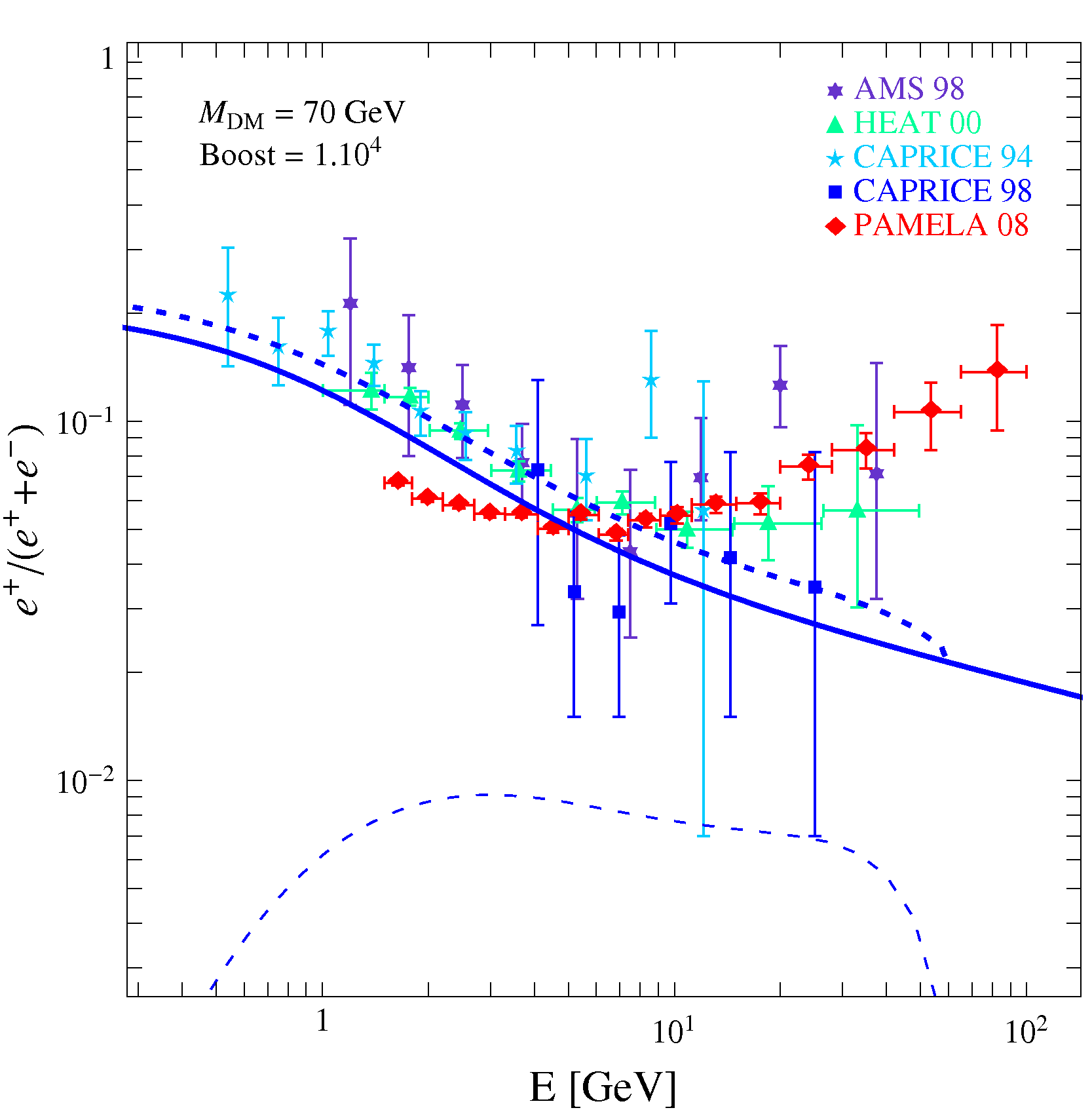}
\includegraphics[width=5.6cm]{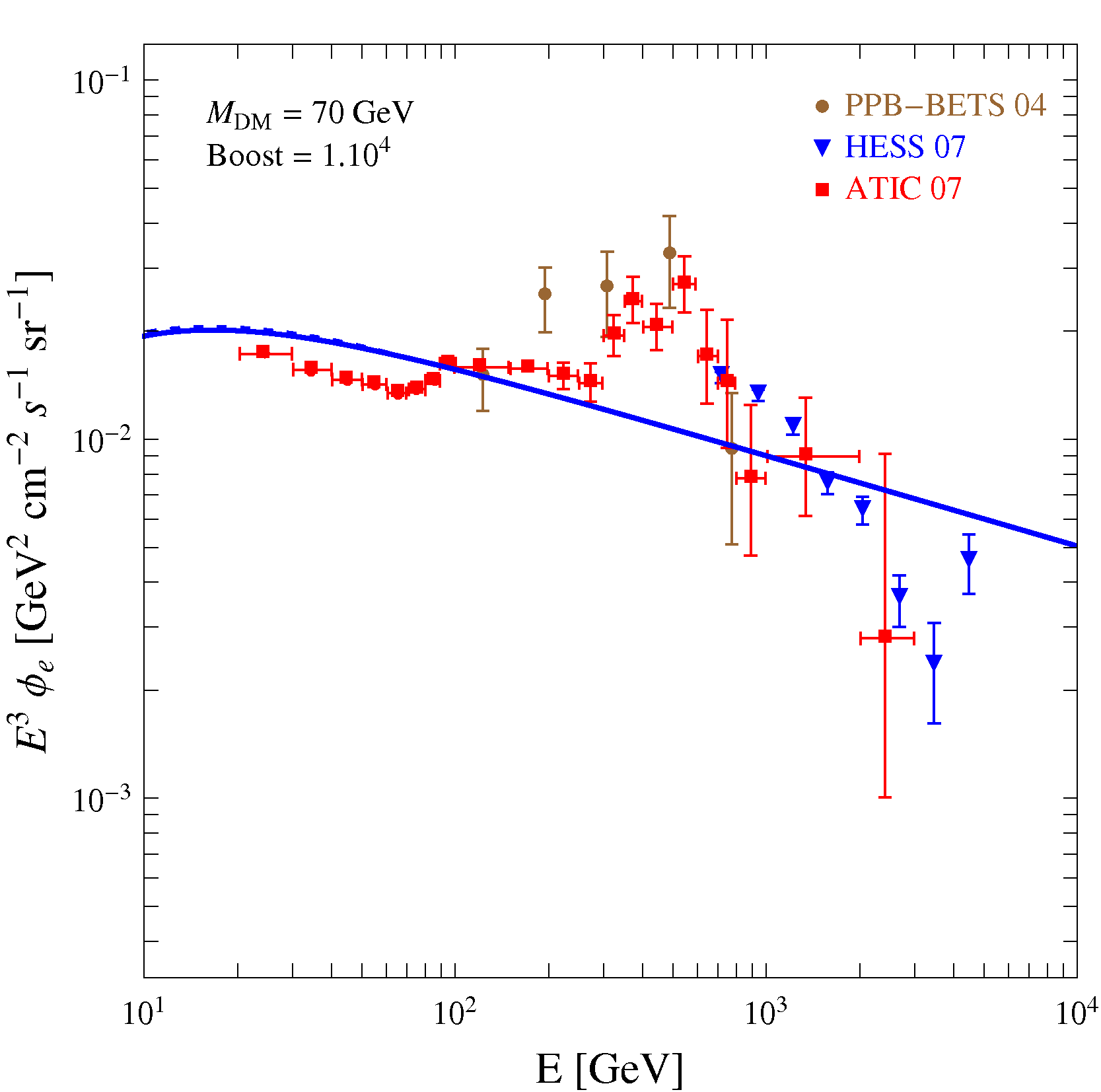}
\includegraphics[width=5.45cm]{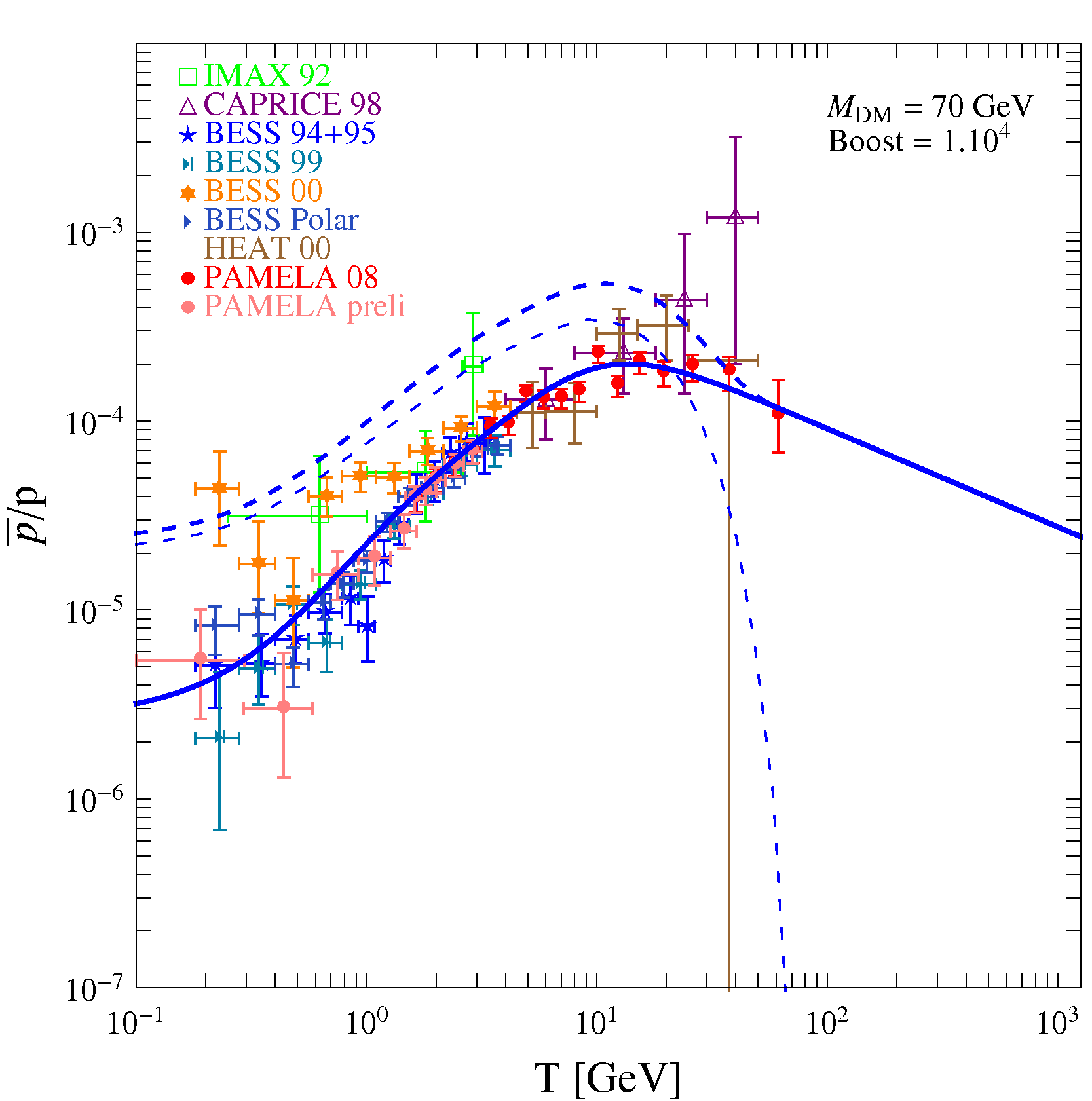}
\caption{Same as Fig.~\ref{fig:lowmass} but for a Middle Mass candidate ($M_{DM}=70$\,GeV, BF = 10$^4$) with one-loop $\gamma Z$ contribution.}
\label{fig:middlemassZg}
\end{center}
\end{figure}

All these fluxes have been computed for $M_h = 120$ GeV. However it is
of interest to envision larger Higgs masses, since this is one of the
historical motivations for the IDM. In \cite{Barbieri:2006dq}, it has been shown
that Higgs as heavy as $M_h \sim 500$ GeV may be consistent with LEP precision
measurements. Although one expects that the annihilation cross section is  small for such a large Higgs mass,
it has been shown in \cite{Gustafsson:2007pc} that the WMAP abundance may be
reached using co-annihilation. In these scenarios, loop effects may be large and the dark matter candidate could have a large branching ratio into
$\gamma$ pairs and into $\gamma Z$. Such gamma rays could be a promising signal for the GLAST/FERMI satellite. 
Although these properties require some
amount of fine tuning, the prediction of significant gamma ray lines is  specific to the
IDM scalar dark matter candidate. Given that a $Z$ may decay into lepton pairs, it is interesting to investigate
whether loop effects may also  enhance the production of positrons.
To address this question, we have considered, as an instance, the benchmark Model I of
\cite{Gustafsson:2007pc}, and have simply taken into account the
positrons coming from $Z$ decay. At one-loop, there is also a direct
contribution with $\bar l l \gamma$ annihilation
\cite{Bergstrom:2008gr}, but we have
neglected it for our estimate. This approximation is {\em a posteriori}
legitimated by the fact that we conclude from  Fig.\ref{fig:middlemassZg} that the signal in positrons from loop effects is too small to be observable,
unless the boost factor is indeed very large, BF $\sim 10^4$. Furthermore,
although the shape of the spectrum is slightly harder for $E \lsim M_{DM}$ than it is at three level, loop effects may not reproduce the observed excesses and, at best,  we could put a limit on the BF.


\subsection{High mass range}

\begin{figure}[h!]
\begin{center}
\includegraphics[width=5.5cm]{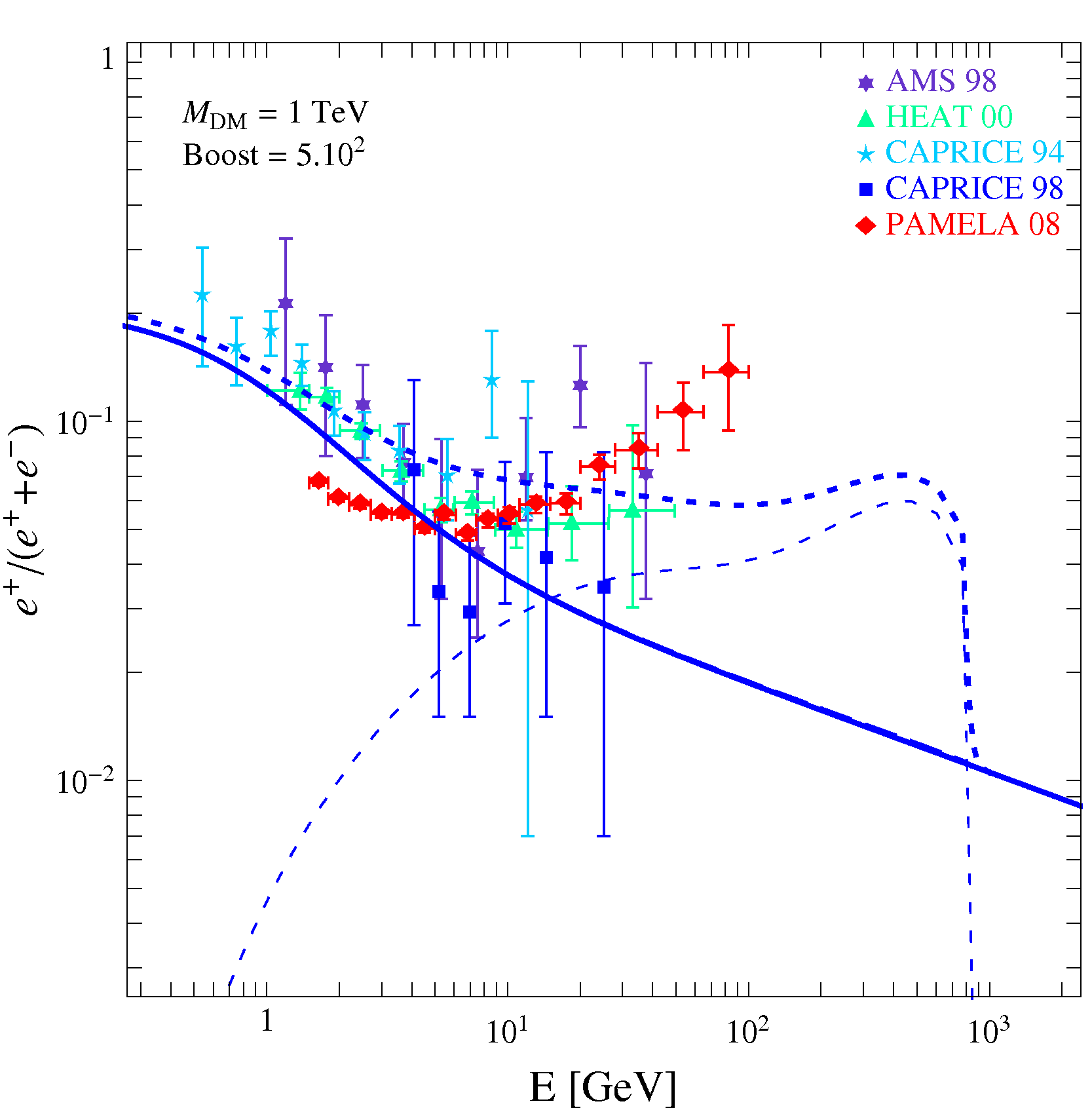}
\includegraphics[width=5.6cm]{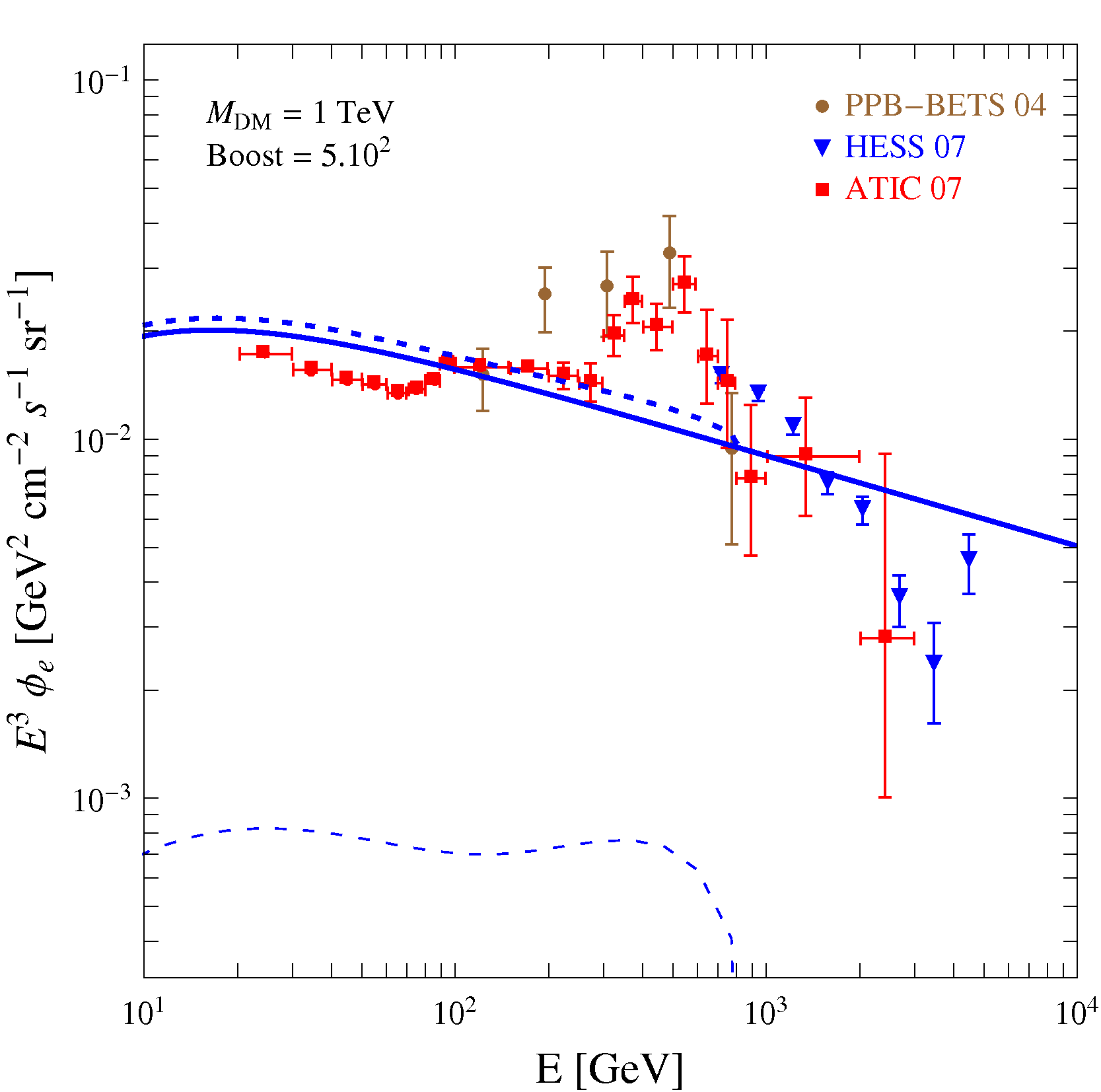}
\includegraphics[width=5.45cm]{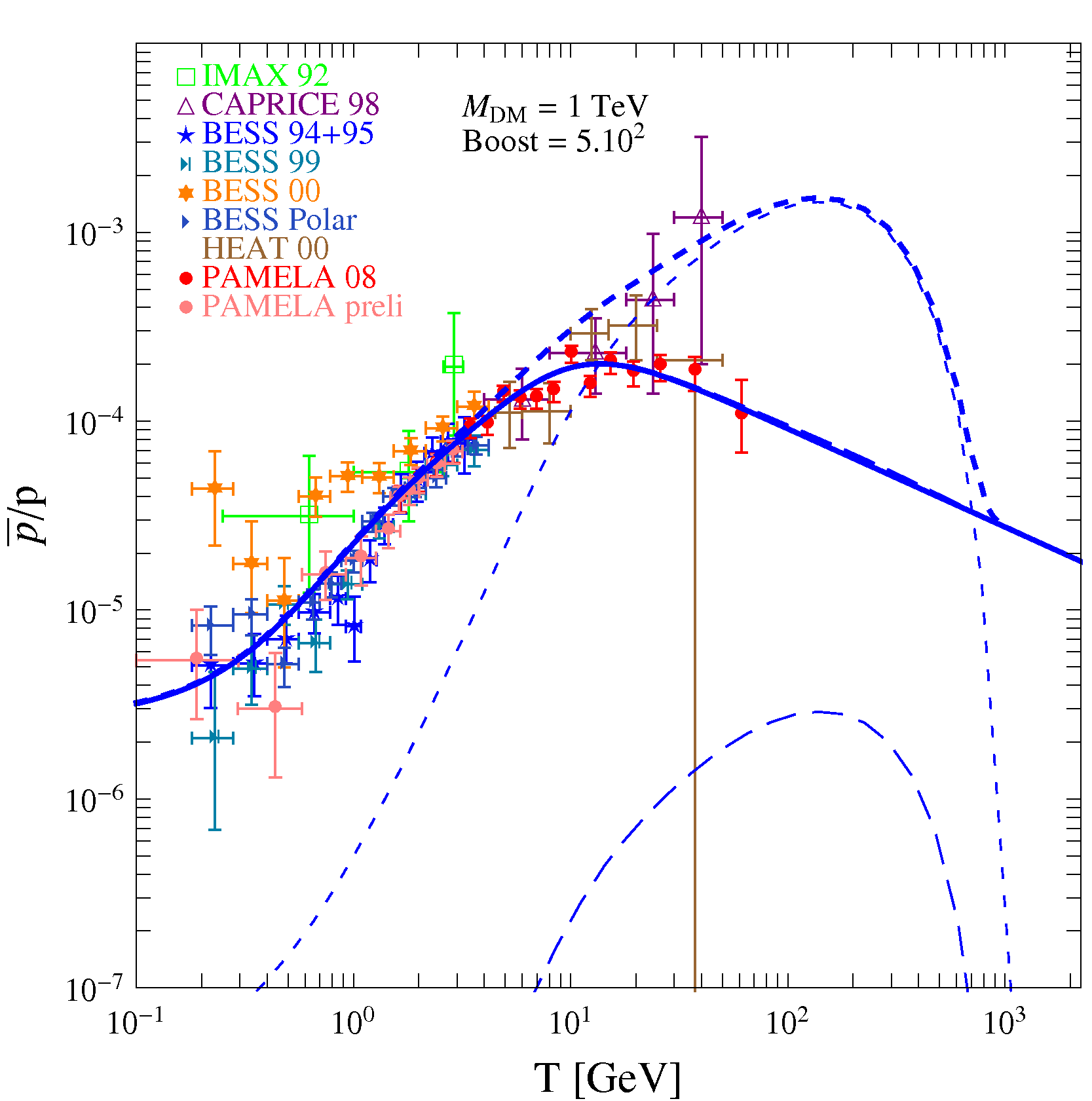}
\caption{Same as Fig.~\ref{fig:lowmass} but for a High Mass candidate ($M_{DM}=1$\,TeV, BF = 5.10$^2$).}
\label{fig:highmass}
\end{center}
\end{figure}

\begin{figure}[h!]
\begin{center}
\includegraphics[width=5.5cm]{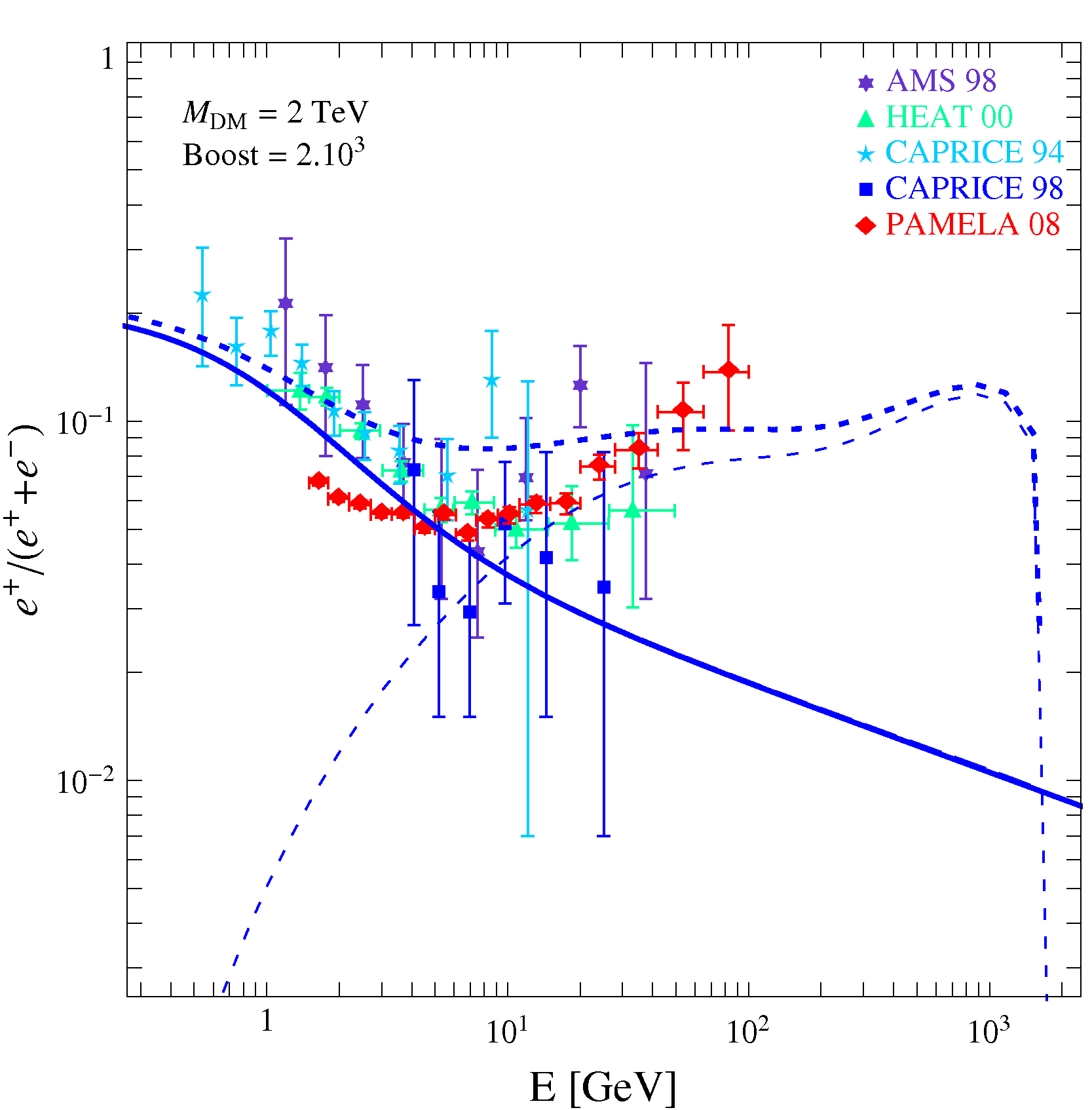}
\includegraphics[width=5.6cm]{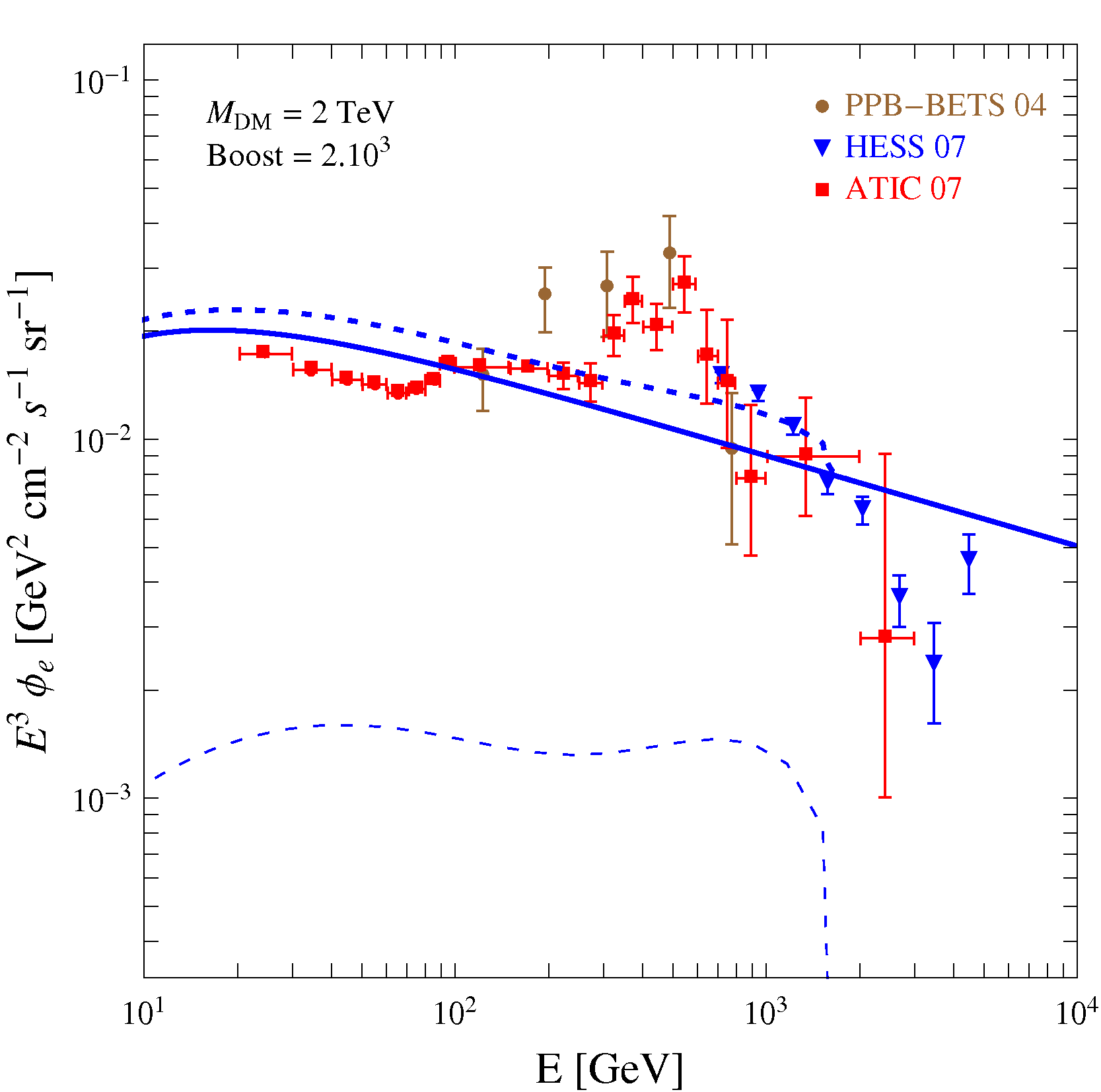}
\includegraphics[width=5.45cm]{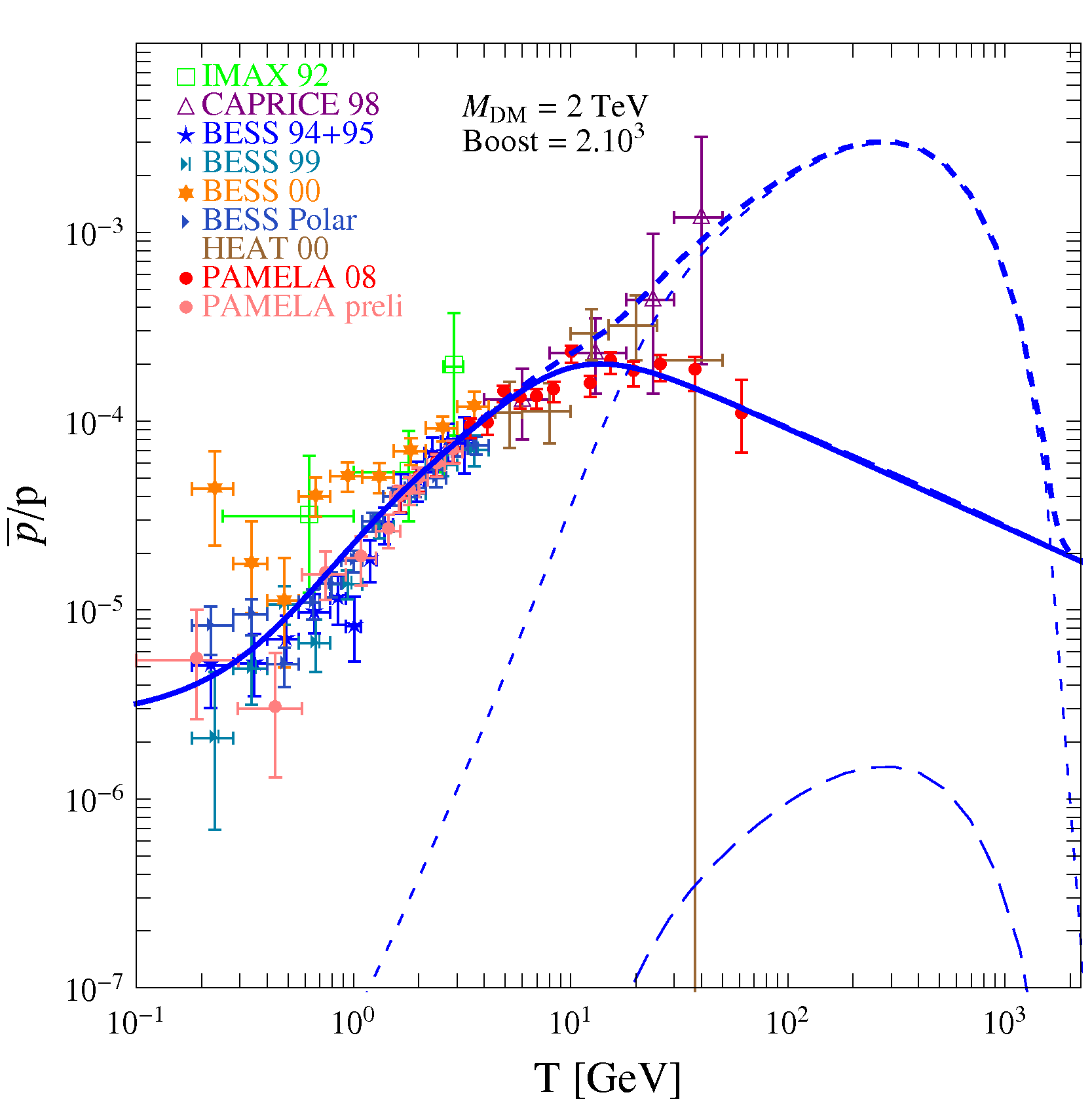}
\caption{Same as Fig.~\ref{fig:lowmass} but for a High Mass candidate ($M_{DM}=2$\,TeV, BF = 2.$10^3$).}
\label{fig:highmass2}
\end{center}
\end{figure}

\begin{figure}[h!]
\begin{center}
\includegraphics[width=5.5cm]{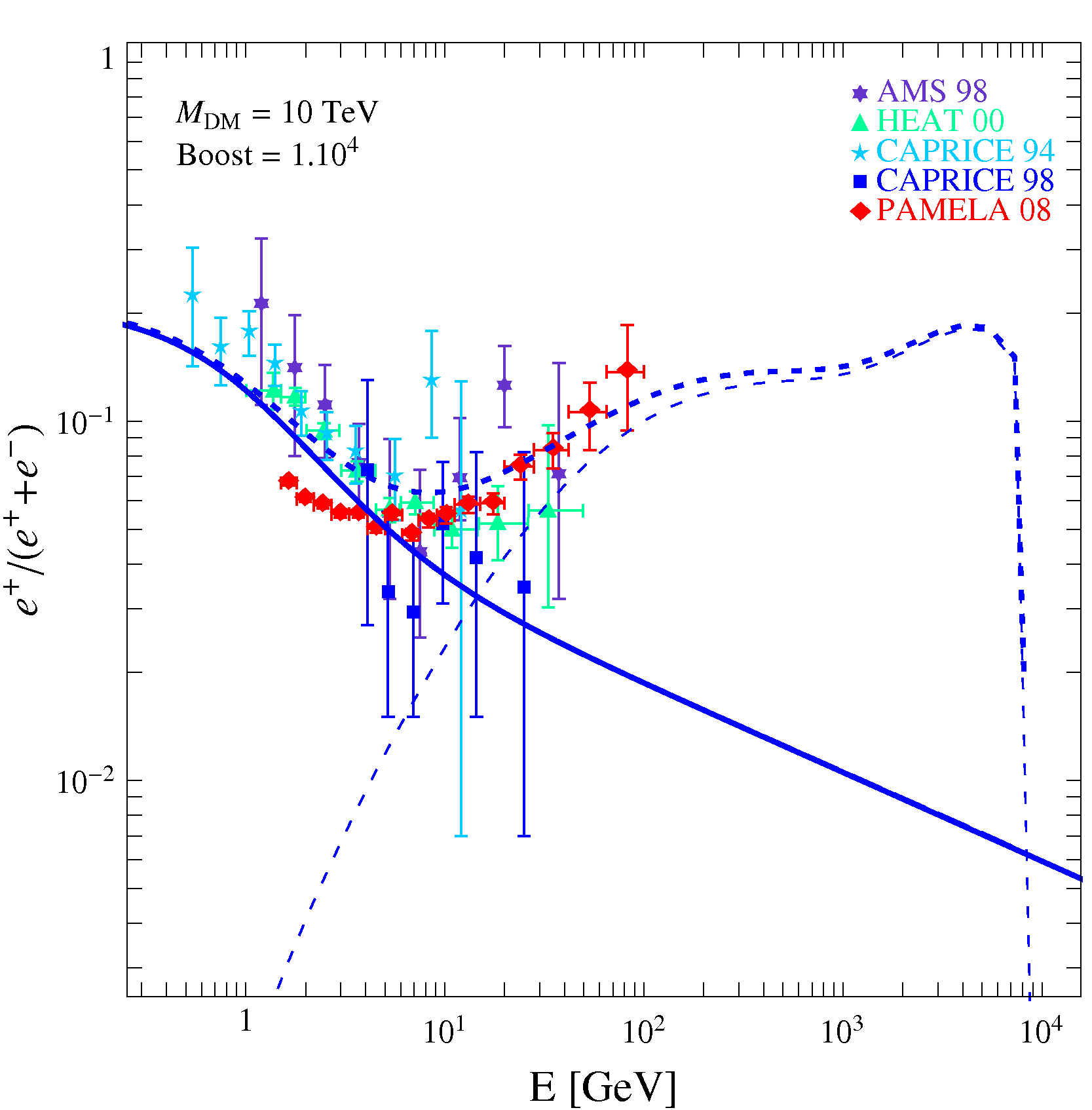}
\includegraphics[width=5.6cm]{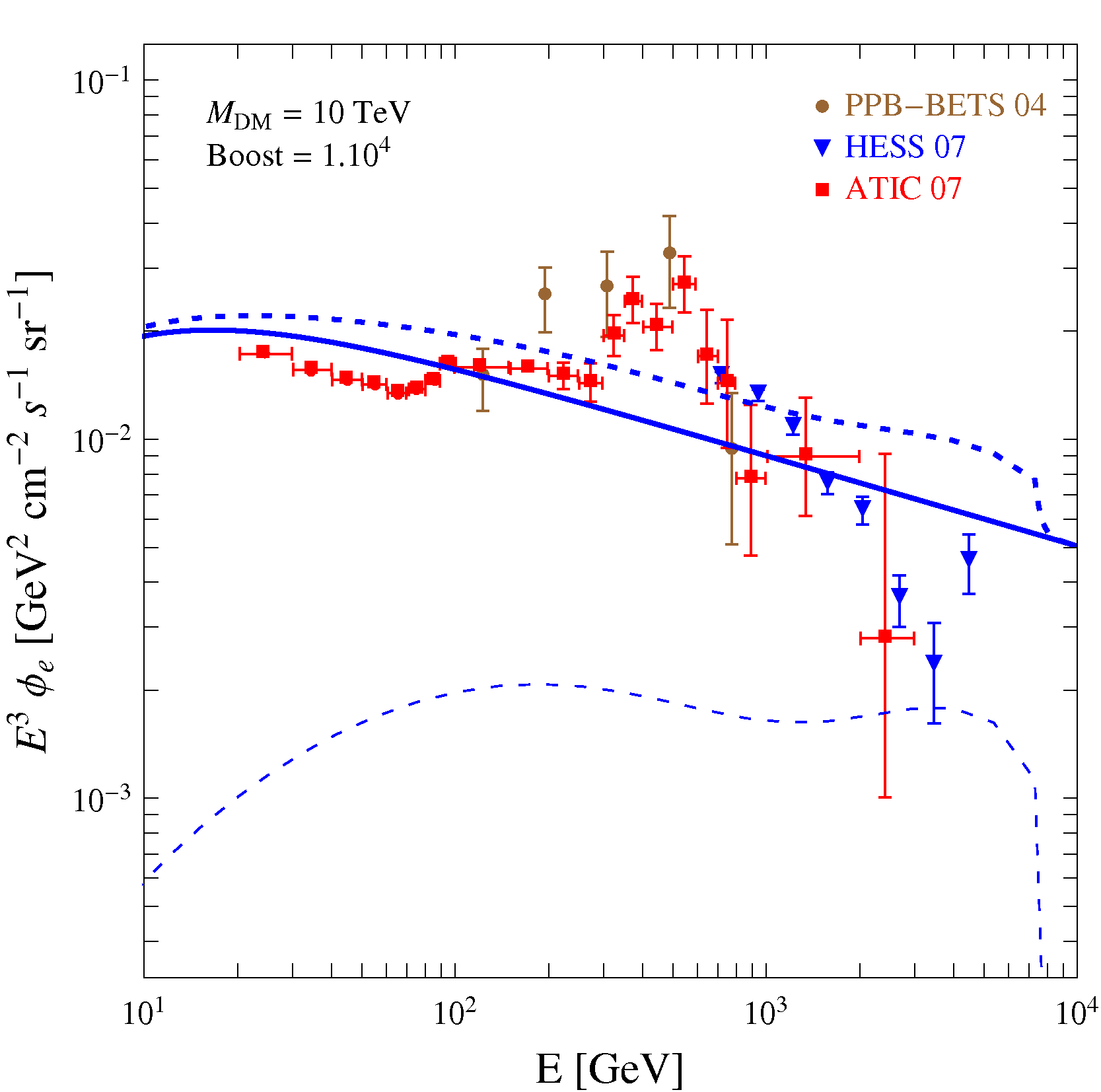}
\includegraphics[width=5.45cm]{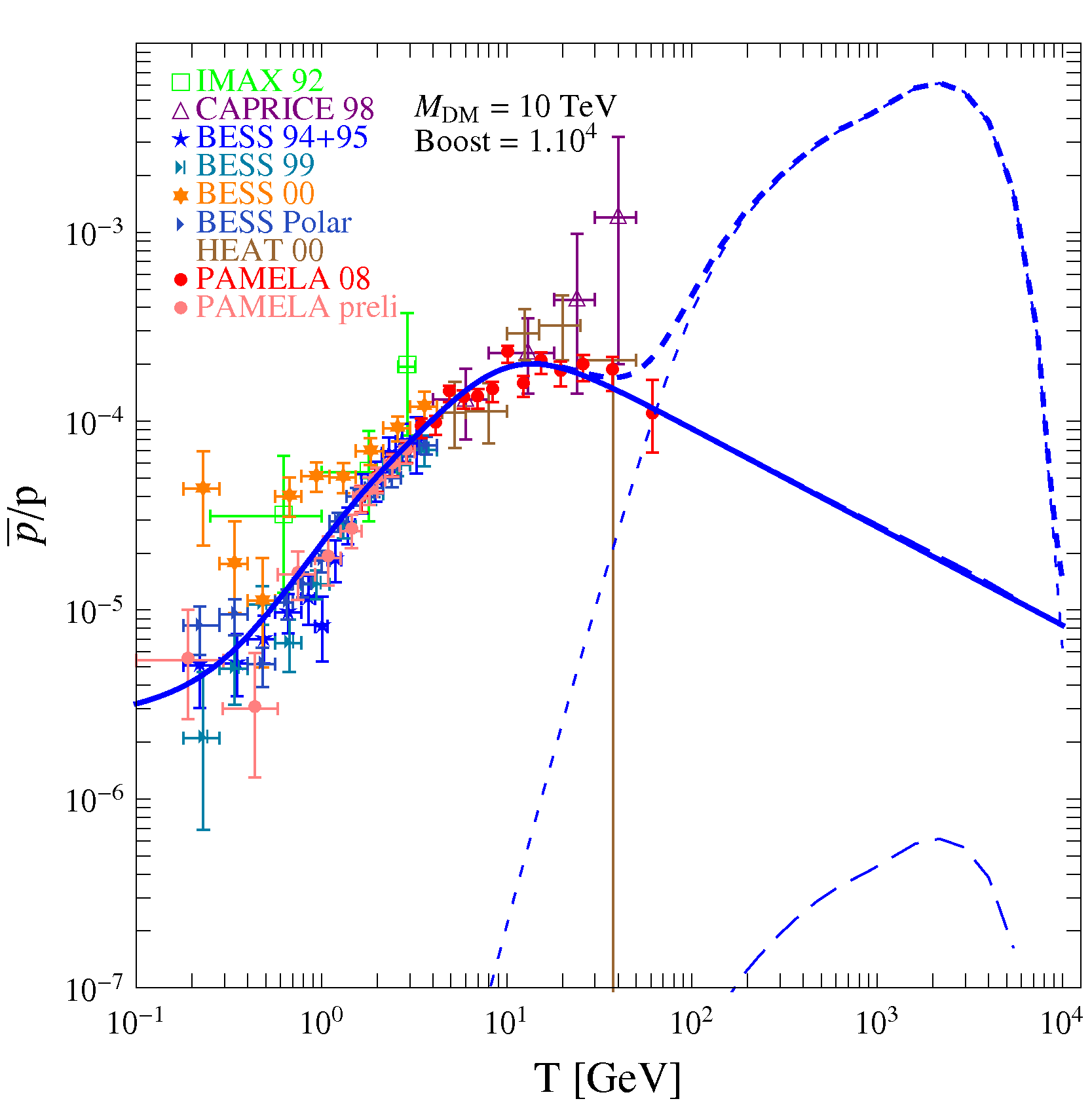}
\caption{Same as Fig.~\ref{fig:lowmass} but for the High Mass regime ($M_{DM}=10$\,TeV, BF = 10$^4$).}
\label{fig:highmass3}
\end{center}
\end{figure}

For a heavier IDM dark matter candidate, annihilations into weak gauge bosons become dominant. Specifically, in the IDM, we may consider
dark matter candidates in the mass range
$$
500 \mbox{\rm\, GeV} \lsim M_{DM} \lsim 15 \mbox{\rm\, TeV},
$$
where the upper bound comes from unitarity limit. In \cite{LopezHonorez:2006gr}, it has been shown that the WMAP cosmic abundance requires the bare mass $\mu_2$ to be close to $M_{H_0}$, and thus, for most the parameter space, annihilations take place essentially into $W^+W^-$ and $Z$ pairs, just like for Minimal Dark Matter candidates \cite{Cirelli:2005uq}. The Figures (\ref{fig:highmass}), (\ref{fig:highmass2}) and (\ref{fig:highmass3}) illustrate the typical signals one may expect from IDM candidates with masses $M_{DM}=$ 1, 2 and 10 TeV. Note that, in these cases, we have adopted an isothermal profile for the distribution of dark matter in the Galaxy, so as not to contradict constraints coming from synchrotron radiation emission by dark matter annihilation products\cite{Bertone:2008xr}.  
The ATIC balloon experiment shows an excess between 300 and 800 GeV which is consistent with the first data points of HESS. It is  tempting to associate those signatures as being due to annihilation of a dark matter particle with a mass around TeV. For this to work requires  a large boost factor BF $\sim 10^3$, however, as for Minimal Dark matter, this boost may come from a combination of astrophysics and particle physics effects, the latter in the form of Sommerfeld enhancement. This being assumed, the strongest constraint comes again from $\bar p$  data from PAMELA, which require  the candidate to have a small branching ratio into baryons (see {\em e.g.} \cite{Cirelli:2008pk}), or, focusing  on the IDM, to be heavier than 10 TeV (Fig.\ref{fig:highmass3}). Unfortunately, as it is well known, the fit to ATIC/HESS is then quite poor. 

One may envision improving the fit to the data by extending the IDM, for instance  by adding more particles. 
This opens the possibility to have direct annihilation into charged lepton pairs\footnote{The simplest and most natural extension of the IDM is to add heavy singlet neutrinos with odd parity \cite{Ma:2006km} but this model does not lead to charged lepton pair annihilation channels. A possibly interesting alternative might be to embed the IDM in a Universal Extra Dimension (UED) model (see for instance \cite{Hooper:2007qk}), in which case annihilation in charged lepton pairs is allowed.}. However, for a scalar DM candidate like in the IDM, these processes would in general be  p-wave or chirality ($\propto m_l^2$) suppressed, and thus not very useful to explain the excesses. More sophisticated extensions may however be considered, as in \cite{Cao:2009yy}, which do not have such a limitation. 

\section{Conclusion}

We have confronted the IDM scalar dark matter candidate to the recent data on antimatter in cosmic rays. Not surprisingly, the fits to data are generically poor compared to those of the recent models that have been designed to explain the PAMELA and ATIC excesses. 
If one insists on explaining PAMELA and ATIC, the most promising possibility offered by the IDM is that of a very heavy DM candidate, with $M_{DM} \gsim 10$ TeV so as to evade the constraints on the anti-proton flux. In this regime, the predictions of the IDM are similar to those of a Minimal Dark Matter candidate, and the price to  pay is the same, as both require  a large boost factor to enhance the fluxes. The Sommerfeld effect, which is generically speaking relevant for a heavy dark matter candidate, may help, but still, some astrophysical boost factor would  be required. 

In our opinion, a most interesting aspect of the IDM is the large flux of positrons, anti-protons, and anti-deuterons that are predicted to be produced in the annihilation of a light WIMP,  with $M_{DM} \lsim 10$ GeV. This is due to two simple features.  The first one is that the annihilation rate is large, dominated by Higgs exchange, in order to reach the WMAP abundance. The second feature is that the flux is proportional to the number density squared, which is large for a light candidate. 
One issue in confronting  a light WIMP candidate to observations is that the low energy cosmic rays are very sensitive to solar modulation, a phenomenon which is still poorly understood.
If progress could be made on this issue, our impression is that useful constraints might be put on the properties of light dark matter candidates, with $M_{DM}\lsim 10$ GeV, a category that actually encompasses many  models of dark matter. In the meantime, a prediction that is not obscured by solar modulation is that the flux of anti-deuterons produced by a light IDM dark matter candidate is predicted to be above the expected AMS-02 and GAPS experiments sensitivities
.

\section*{Acknowledgments} 
The work of G. Vertongen and M.H.G. Tytgat is supported by the FNRS and Belgian Federal Science Policy (IAP VI/11).





\end{document}